\documentclass[pss]{wiley2sp}
\usepackage{amsmath}
\usepackage{bm}
\usepackage{w-greek}

\graphicspath{{figs/}
      }

\newcommand{\degr}{$^\circ$}
\newcommand{\etal} {\textit{et al.}}
\newcommand{\ie} {\textit{i.e.}}
\newcommand{\eg} {\textit{e.g.}}

\newcommand{\sapp} {Al$_2$O$_3$}
\newcommand{\laal} {LaAlO$_3$}

\begin{document}

\title{Non-DMS related ferromagnetism in transition metal doped zinc oxide}

\titlerunning{Non-DMS related ferromagnetism}

\author{%
Kay Potzger\textsuperscript{\textsf{\bfseries     1,\Ast}},
Shengqiang Zhou\textsuperscript{\textsf{\bfseries      1}}}

\authorrunning{K. Potzger, et al.}

\mail{e-mail \textsf{k.potzger@fzd.de}, Phone +49-351-2603148, Fax
+xx-xx-xxx}

\institute{%
\textsuperscript{1}Institute for Ion Beam Physics and Materials
Research, Forschungszentrum Dresden-Rossendorf, Bautzner
Landstra{\ss}e 128, 01328 Dresden}
\received{XXXX, revised XXXX, accepted XXXX} 
\published{XXXX} 


\abstract{%

\abstcol{We review pitfalls in recent efforts to make a
conventional semiconductor, namely ZnO, ferromagnetic by means of
doping with transition metal ions. Since the solubility of those
elements is rather low, formation of secondary phases and the
creation of defects upon low temperature processing can lead to
unwanted magnetic effects. Among others, ion implantation is a
method of doping, which is highly suited for the investigation of
}{those effects. By focussing mainly on Fe, Co or Ni implanted ZnO
single crystals we show that there are manifold sources for
ferromagnetism in this material which can easily be confused with
the formation of a ferromagnetic diluted magnetic semiconductor
(DMS). We will focus on metallic as well as oxide precipitates and
the difficulties of their identification. }}

\maketitle     

\section{Introduction: A little history of (diluted) magnetic semiconductors }
Magnetic semiconductors are recently considered as one of the main
building blocks for future replacement of conventional electronics
by spin-electronics or spintronics \cite{wolf01}. The latter is a
field of materials research born from the need of higher speed,
lower power consumption and smaller integration densities of
computational devices. A model device of spintronics was proposed
already in the year 1990, i.e. the Datta-Das field effect
transistor \cite{datta90}. The Datta-Das transistor allows the
modulation of the spin-polarized electron current flowing through
a two dimensional electron gas. In contrast to other devices, due
to application of a gate voltage, the spin direction of the
electrons and thus the magneto-resistance between source and drain
is altered. For creation and detection of spin polarized electron,
Datta and Das proposed ferromagnetic iron contacts. The efficiency
of spin injection from metals, as well as other materials, into
semiconductors was recently described by A. M. Bratkovsky
\cite{bratkovsky}. The main obstacle is the large conductivity
mismatch between the ferromagnetic metal and the semiconductor.
The latter strongly reduces the spin polarization in the
semiconductor down to below 0.1\% \cite{PhysRevB.62.R4790} even
for a spin polarization of $<$ 99\% in the metal contact. Thus,
low conductivity mismatch at 100\% spin polarization is required.
Among other solutions like tunnel barriers or layers between the
metal and the semiconductor, the use of ferromagnetic
semiconductors as source contacts is proposed. A well known
semiconductor with spin polarized charge carriers is EuO. EuO has
a band gap of 1.12 eV and a Curie temperature of 69 K. The spin
splitting of the conduction band amounts to 0.6 eV. By means of
doping, the conductivity of EuO can be tuned. These basic
properties are shortly reviewed by A. Schmehl et al.
\cite{schmehl}, who showed that the spin polarization of charge
carriers in epitaxial EuO exceeds 90\%.  Moreover, a
heterostructure based on EuO grown on GaN or Si was realized. The
low Curie temperature of EuO, however, makes it inappropriate for
everyday electronics. Therefore, efforts for increasing its Curie
temperature as well as mining for alternative spin-injectors are
undertaken. One alternative is considered to be a ferromagnetic
diluted magnetic semiconductor (DMS). DMS materials have been
investigated already about 30 years ago \cite{furdyna:R29}. The
investigations concentrated mainly on II-VI semiconductors like
CdSe or CdTe doped with up to 45\% of Mn$^{2+}$. These compounds,
although paramagnetic, exhibit extraordinary properties at low
temperatures and/or large magnetic fields. Such effects as giant
Faraday rotation, bound magnetic polaron, giant Zeeman splitting
of the electronic (band and impurity) levels originate from the
magnetic exchange interaction between the sp electrons of the
semiconductor host and the localized d-electrons of the Mn
impurity. The exchange integral was actually determined to be
larger for p-d than for s-d interaction. Thin film DMS based on Mn
doped III-V semiconductors have been grown successfully in the
1990 ies by low temperature molecular beam epitaxy, i.e. {InMnAs}
\cite{munekataInMnAs} and GaMnAs \cite{ohnoGaMnAs}. The field of
DMS was greatly pushed in the year 2000 when Ga$_{1-x}$Mn$_x$As
was, both experimentally and theoretically, found to be
ferromagnetic up to 110 K. The mechanism of the ferromagnetic
interaction is based on the exceptional case that Mn doping
introduces shallow acceptors and thus holes into the valence band
of the GaAs that are antiferromagnetically coupled to the local Mn
moments. T. Dietl calculated the Curie temperature of different
p-type semiconductors doped with Mn ions \cite{dietl00}. He found
that it is proportional to the concentration of the impurities and
to the squire root of the p-type charge carrier concentration.
Recently, the Curie temperature of Ga$_{1-x}$Mn$_x$As could be
increased up to 173 K by means of increasing its crystalline
quality and the number of Mn impurities up to 6.8\%
\cite{jungwirth:165204}. Room-temperature ferromagnetism has been
predicted for 5\% Mn doped p-type ZnO or GaN. Note that the
predictions have been criticised recently \cite{sato05}. Later,
the theory was extended to n-type material with V, Cr, Fe, Co or
Ni dopants \cite{sato_ZnO}. The Curie temperature for those cases
is predicted to scale with the square root of the impurity
concentration. Note that for selected TM there are sometimes
exceptions from the equilibrium solubility limits, i.e. wurtzite
Zn$_{1-x}$TM$_x$O with x$>$5\% can be formed. One example is
Rinmans green (well known since 1780), a pigment basing on
Zn$_{1-x}$Co$_x$O, created from firing of a mixture of ZnO and CoO
powders. Recent experiments on the creation of Zn$_{1-x}$Co$_x$O
thin films with controlled transport properties has been performed
by pulsed laser deposition, but especially high quality films only
show paramagnetism
\cite{xu:013904,xu:063918,liu:154101,xu:205342,ney:157201}.
Besides ferromagnetic order induced by indirect d-d exchange,
defect mediated coupling of the local d-electrons was recently
considered. J. M. D. Coey \etal~developed a model where triplet
molecular orbitals form an impurity band. If narrow enough, the
impurity band can split spontaneously. On the other hand it can
also indirectly couple localized electrons (if present)
\cite{coey05}. As point defects for mediation of ferromagnetic
coupling, e.g. between dispersed Co ions, usually oxygen vacancies
or Zn interstitials are considered
\cite{hsu:242507,kittilstved:037203}. Those are believed to
provide n-type charge carriers. In ref. \cite{kittilstved:037203},
Zn interstitials have been introduced and removed in a controlled
way, altering the magnetic properties of Co doped ZnO. On the
other hand Zn vacancies are also proposed to mediated TM
d-electrons by other groups
\cite{weyer:113915,yan:062113,karmakar:144404,xu:092503}. In Ref.
\cite{xu:092503} the authors identified a deep acceptor trap due
to Zn vacancies and this finding suggests the possibility of
hole-mediated ferromagnetism in ZnO. Kittilstved et al. recently
pointed out the important role of uncompensated p-type dopants for
the creation of ferromagnetic Mn-diluted ZnO \cite{kittilstved05}.
Further evidence for defects enhancing ferromagnetism in Co doped
ZnO is given in Refs. \cite{songJPC,pan08}. The common feature
about the above mentioned research is the indirect magnetic
coupling of localized d-moments leading to spin-polarized
currents. By creating various layered architectures basing on TM,
but mainly Co doped ZnO, magneto-resistance effects have been
observed which proof the creation of such currents from the
exchange interaction between localized and itinerant charge
carriers \cite{song:172109,song:042106,xu:076601,ramachandran}.
Especially the combination of magneto-optical measurements and
anomalous Hall effect is mentioned to confirm this effect
\cite{behan:047206}. In our paper we would like to point out
sources for non-intrinsic, i.e. non DMS related ferromagnetic
properties in order to sharpen the eye against false
interpretation.

\section{Analysis techniques}

In table \ref{tab:method}, we shortly give an overview of analysis
techniques we used during our research and point out relevant
information provided by the methods. The main obstacle about
choosing the right analysis technique arises from the fact that
magnetic signals detected in transition metal doped ZnO usually
are small so that they are sometimes only detectable by SQUID
magnetometry. This technique, however, is an integral method.
There are several modi for data acquisition like magnetization vs.
field, temperature or time measurements giving deeper insight
about the character of the magnetic ordering. Nevertheless there
are several pitfalls about using SQUID recently collected in ref.
\cite{ney2008}. Therefore it is desired to use element specific
analysis methods in order to correlate the magnetic properties
with the element responsible for them. One of the most basic
methods in investigating of TM impurities as well as intrinsic
defects with unpaired spins in ZnO is electron paramagnetic
resonance (EPR). As technique, EPR spectroscopy is based on a
detection of resonance absorption of electromagnetic energy
corresponding to transitions between electron-spin energy levels
split by internal effects (crystal field anisotropy, exchange
interactions, spin-orbital effects) and/or by an applied magnetic
field.  Mn, Fe, and Cu impurities as well as single crystals have
been investigated by EPR already a few decades ago. For Cu doped
ZnO \cite{PhysRev.132.1559}, a 2+ state indicating Zn substitution
was found. The main conclusion is the extension of the Cu t$_2$
acceptor wave function towards the nearest neighbours in the
crystal. Mn impurities in ZnO have been observed by EPR already in
1958\cite{PhysRev.112.1058}. One of the most interesting 3d
impurities in ZnO is Fe, since it occurs, although substitutional,
in a 3+ oxidation state \cite{PhysRev.126.952}. The general
tendency of Fe to develop the Fe$^{3+}$ state in II-VI materials
can be understood from the stable 3d$^5$ configuration achieved
(similar to Mn$^{2+}$ in III-V compounds). Recently, the mechanism
of charge compensation of the Fe$^{3+}$ centers in hydrothermal
ZnO single crystals was investigated \cite{azamata}. It was
proposed that nearest neighbour vacancies act as compensators. We
would like to add that in such case the Fe$^{3+}$ center would
also stabilize such vacancy. Using EPR, it is also possible to
investigate paramagnetic properties of intrinsic defects in ZnO
like zinc \cite{galland} or oxygen vacancies
\cite{smith70,leiter01,vlasenko:125210}. Usually electron
irradiation is used for defect formation.

Another element specific analysis technique is M\"{o}ssbauer
spectroscopy. Although spin-split sublevels of the nucleus are
investigated, properties of the local electrons like charge state
or magnetic ordering can be made. The parameters investigated are
the isomer shift, the quadrupole and the magnetic hyperfine
splitting at the nucleus. Early investigations on solid solutions
of $^{57}$Fe, Zn and O mainly revealed oxide spinel secondary
phases to be responsible for magnetic sextets observed
\cite{dobbon70,tanaka94}. There are other methods basing on
radioactive probes which definitely have the potential to give
more insight into the origin of ferromagnetic properties in TM
doped ZnO. One is perturbed angular correlation spectroscopy (PAC)
mainly using $^{111}$In/$^{111}$Cd as a probe. Since it is easily
incorporated into Zn substitutional lattice sites \cite{deubler92}
and does not exhibit local magnetic moments, it can serve as a
probe for the spin polarization of the s electrons polarized by
ferromagnetically aligned d-electrons.

A method for element specific analysis of the magnetic properties
of TM doped ZnO is x-ray magnetic circular dichroism (XMCD)
available a synchrotron radiation sources. The technique bases on
the dichroic absorption of left- and right circular synchrotron
light at a fixed magnetic field or at a fixed helicity while the
field direction is reversed. Until now, there are no reports on
low-field dichroic signals of TM doped ZnO based DMS which are not
doubtlessly related to ferromagnetic inclusions. In contrast, the
investigation of paramagnetic Co doped ZnO with perfect Wurtzite
structure became very popular in recent years. One report
combining high quality level preparation as well as analysis
showed purely paramagnetic properties with a magnetization of 4.8
$\mu_B$ per Co ion. There is evidence for antiferoomagnetic
coupling. For TM doped ZnO containing metallic ferromagnetic
secondary phases, XMCD is especially valuable because the
metallic/ionic character can be distinguished while simultaneously
the magnetization can be determined \cite{ney:157201}.

\begin{table}
  \caption{Analysis methods used during our research.}
  \renewcommand{\thempfootnote}{\fnsymbol{footnote}}
  \begin{minipage}{\hsize}
  \begin{tabular}[htbp]{p{1.5cm}p{1.2cm}p{4.3cm}}
    \hline
    Method & Device & Information \\
    \hline
    SQUID\footnotemark[1] & QD MPMS XL 7T  & - Exact determination of moment \newline
                              - Purely integral method \\
    \hline
    CEMS\footnotemark[2]   &                & - Hyperfine parameters at $^{57}$Fe nucleus \newline
                              - Isotope specific \newline
                              - Large information depth in crystal  \\
    \hline
    XMCD\footnotemark[3]   &  Synchrotron   & - Spin-and orbital moment \newline
                              - Surface and bulk sensitive \newline
                              - All advantages of XAS  \\
    \hline
    XAS\footnotemark[4]   &   Synchrotron   & - Charge states \newline
                              - Crystal field splitting \newline
                              - Element specific  \\
    \hline
    RBS/C\footnotemark[5] &   Ion beam lab  & - Absolute impurity concentration \newline
                              - Impurity lattice position \newline
                              - Lattice disorder \\
    \hline
    XRD\footnotemark[6] &   Siemens D5005   & - Identification of second crystalline phases \newline
                              - Crystalline lattice quality \\
    \hline
    SR-XRD\footnotemark[7] &   Synchrotron  & - Larger sensitivity than lab-XRD \\

    \hline
    HR-XRD\footnotemark[8] &  Siemens D5005 & -HR reciprocal space mapping (RSM) for strain or imperfections
    determination \\
    \hline
    TEM\footnotemark[9] &   FEI Titan   & - High resolution real space method for second phase or defect characterization \newline
                              - Non-integral method \\
    \hline
    PAS\footnotemark[10] &                 & - Characterization of open volume defects \newline
                              - Insensitive to negatively charged
                              defects\\
    \hline
    AFM/MFM\footnotemark[11] &  Veeco/DI Multi-mode   & - Characterization of surface structure \newline
                              - Simultaneous recording of magnetic force
                              gradients \\
    \hline
  \end{tabular}
  \label{tab:method}
  \footnotemark[1] {superconducting quantum interference
  device}\newline
  \footnotemark[2] {conversion electron M\"{o}{\ss}bauer spectroscopy}\newline
  \footnotemark[3] {x-ray magnetic circular dichroism}\newline
  \footnotemark[4] {x-ray absorption spectroscopy}\newline
  \footnotemark[5] {Rutherford backscattering/channelling}\newline
   \footnotemark[6] {x-ray diffraction}\newline
  \footnotemark[7] {synchrotron radiation x-ray
diffraction}\newline
  \footnotemark[8] {high resolution x-ray diffraction}\newline
  \footnotemark[9] {transmission
electron microscopy}\newline
  \footnotemark[10] {positron annihilation spectroscopy}\newline
  \footnotemark[11] {atomic/magnetic force microscopy}
  \end{minipage}
 \end{table}

\section{Unintended ferromagnetism - role of the substrates}

\begin{figure} \center
\includegraphics*[scale=0.7]{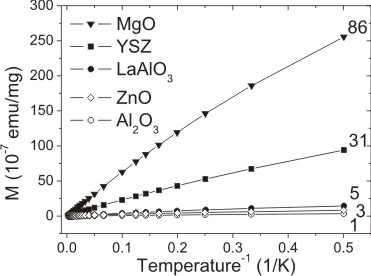}
\caption{%
M-1/T dependence for different substrates at an applied magnetic
field of 10 kOe. The curves can be fairly fitted by a Brillouin
function using a quantum number J of 2.5 and a g-factor of 2 (not
shown). These parameters correspond to Fe$^{3+}$ ions. The values
on the right side times 10$^{13}$ indicate the number of
contaminating Fe$^{3+}$ ions per mg substrate material.}
\label{virgin_sample}
\end{figure}

Ferromagnetic hysteresis and irreversibility of the temperature
dependence in TM doped ZnO might have other sources than the
formation of a ferromagnetic DMS. If the signals are small, one
has especially to consider the role of the substrates. Typical
substrates for the growth of thin films of ZnO or related
materials are typically \sapp, \laal, YSZ, MgO or ZnO. The
commercial sources for single crystalline oxide substrates
nowadays are still limited. We mainly used substrates purchased
from CRYSTEC, CRYSTAL or MATECK (Germany). Table
\ref{tab:as_purch} contains the typical preparation methods for
these materials one has to keep in mind while analysing their
magnetic properties. \sapp~is of high hardness and therefore used
as abrasive, e.g. in ball mills. At high temperatures alumina acts
reducing and therefore is used for the reduction of iron oxide to
metallic iron \cite{paananen}. Therefore, \sapp~is most sensitive
on contamination with metallic Fe during production or handling.
Yttria stabilized ZrO$_2$ (YSZ) has comparable properties to
\sapp. It also acts reducing on iron oxide upon annealing at high
temperatures  \cite{honda:711}. Due to its lower hardness as
compared to steel (7-8 Mohs) ZnO is less sensitive to
contamination from handling with steel tools.

\begin{table}
  \caption{Selected single crystalline substrates used for semiconductor thin film growth along with the growth method and hardness. The latter is given in Mohs scale displaying the ability of a harder material to scratch a softer material. Data was taken from CRYSTEC homepage.}
  \begin{tabular}[htbp]{p{1.0cm}p{5.0cm}p{1.0cm}}
    \hline
    Crystal & Growth method & Hardness (Mohs) \\
    \hline
    \sapp~& Edge Defined Film Fed Growth (EFG)  & 9  \\
    YSZ & Skull melting & 8.7  \\
    LaAlO3 & Verneuill Czochralski & 6.5 \\
    MgO & Arc fusion & 5.5 \\
    ZnO & Hydrothermal & 4.5 \\

    \hline
  \end{tabular}
  \label{tab:as_purch}
\end{table}

\begin{figure} \center
\includegraphics*[scale=0.6]{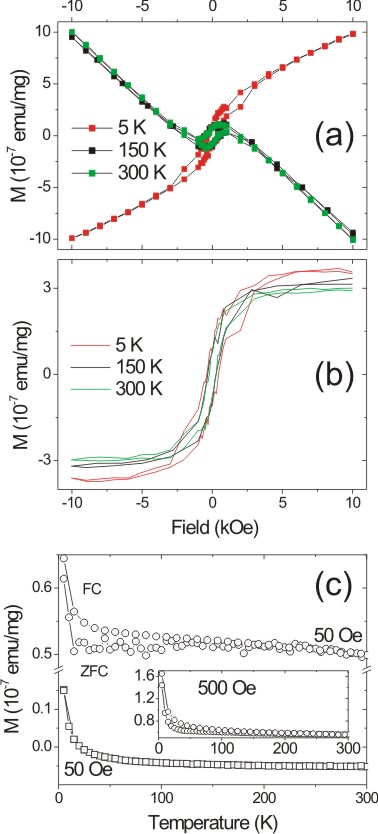}
\caption{%
(a) Magnetization reversal loops for YSZ(001) single crystalline
material exhibiting weak ferromagnetism. For (b) the background of
constant susceptibility was subtracted. Note that the saturation
magnetization is almost independent of the temperature. (c)
displays ZFC/FC magnetization vs. temperature measurements at 50
Oe and 500 Oe (latter inset) for the ferromagneitc YSZ (circles).
For comparison, the ZFC/FC curve of a non magnetic substrate is
displayed at the bottom (squares).} \label{Fig2}
\end{figure}

In addition, note that single crystalline oxide substrates do not
exhibit the high purity of, e.g. Si, due to the different growth
methods. They are contaminated both with spurious elements as well
as defects in the crystalline lattice. The latter range from point
defects (pm scale) over point defect agglomerates (nm scale) to
microscopic defects like grain boundaries. A recent paper on point
defects in ZnO single crystals investigated by PAS (Table
\ref{tab:method}) was recently published by G. Brauer et al.
\cite{brauer07}. The authors showed that the kind of defect in
virgin materials depends on the preparation method. While
pressurized melt grown crystals contain Zn$^+$-O divacancies,
hydrothermal crystals exhibit a defect likely based on a Zn
vacancy. Confirming paramagnetic contamination, Fig.
\ref{virgin_sample} displays the temperature dependence of the
magnetization (M-1/T) of the different substrates at a magnetic
field of 10 kOe applied parallel to the sample surface. The
paramagnetic contamination is largest for MgO substrates and
lowest for \sapp~substrates. Despite the large contamination with
paramagnetic impurities we did not experience any ferromagnetic
signal in none of the substrates with one exception. Fig.
\ref{Fig2} displays the magnetization vs. magnetic field (M-H) a
well as zero-field-cooled/field cooled (ZFC/FC) M-T measurements
for a weakly ferromagnetic YSZ(001) virgin sample. A hysteretic
behaviour in both the M-H and the ZFC-FC curves can clearly be
seen. The saturation magnetization is very low reaching about
3.5$\times10^{-7}$ emu/mg. Since the sample mass was 229 mg, this
value corresponds to a moment of 8$\times10^{-5}$ emu. This value
again corresponds to 4$\times10^{15}$ Fe atoms in metallic state,
if the magnetization of 2.2 ${\mu}_B$ per atom is assumed. The
origin of the ferromagnetic signal, however, is by no means clear.
Due to the low saturation moment as well as its nearly temperature
independence, a contamination with metallic Fe due to abrasion is
likely. Such contamination for similar substrates, i.e. due to
handling with stainless steel tweezers, has been shown earlier
\cite{abraham05}. On the other hand, it should be noted that it is
nearly impossible to really detect such a low amount of metallic
Fe. Therefore, such assumption would be highly speculative and
might lead to false generalization of the origin of weak
ferromagnetic signals in oxides. Instead, the creation of weak
unconventional ferromagnetism by both spurious TM as well as
defects has been stated very recently \cite{coey05}. Paramagnetic
defects in oxides are well know. One of the first observations of
unconventional ferromagnetism in oxides was made by J. M. D. Coey
et al. on HfO$_2$ \cite{coeyhfo}. This material is especially
interesting because it does not contain any d-electrons. The
explanation of the ferrromagnetic properties in HfO$_2$ was given
along with the assumption of triplet ground states or low lying
excited states due to point defects. Another approach is given by
A. Hernando et al. \cite{hernando:052403}. As a first step, the
authors assume broken symmetries such as domain boundaries. These
boundaries lead to orbital states that are sometimes of large
radii. Intra-orbital ferromagnetic spin correlations induce the
alignment of the momenta. All these results are, however, still
under strong debate.

\begin{figure}\center
\includegraphics*[scale=0.8]{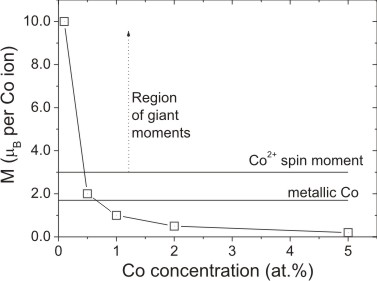}
\caption{%
Fiction magnetic moment per Co$^{2+}$ ion introduced into ZnO at
different concentrations. Assuming a constant ferromagnetic
background arising from the substrate but no contribution from Co,
the sketch would suggest giant magnetic moments to be present at
the Co site. For comparison, the known values for the
magnetization of metallic Co, the Co$^{2+}$ spin moment.}
\label{Fig3}
\end{figure}

Summarizing, three sources are recently believed to provide a weak
unintended ferromagnetic signal in substrates, i.e. contamination
with iron due to handling, spurious elements forming magnetic
clusters and defects. Along with unintended ferromagnetism, a
phrase occurring quite often in literature, i.e. giant magnetic
moments (GMM), has to be discussed. Usually GMM denote a larger
net moment per TM dopant ion than can be expected from its atomic
spin moments. Since the unwanted ferromagnetic signal from the
substrates is not directly related to the TM concentration in the
films, GMM might seemingly occur. There are, however, only a few
experiments where the GMM is assigned directly to the TM dopant,
e.g. from element specific spectroscopy \cite{riegel}. Note that
such GMM are also predicted theoretically for very tiny clusters
containing transition metals \cite{rao02}. If integral
magnetometry is applied and low concentrations of the dopant are
present, a ferromagnetic background from the substrate can easily
account for seemingly GMM in ferromagnetic DMS. Thus the
magnetization, i.e. the moment per dopant atom, appears to vary
along with the dopant concentration, although the TM dopant does
not provide any ferromagnetic contribution. Fig. \ref{Fig3} shows
such a fiction case ZnO doped with different concentrations of Co.
At higher concentrations the magnetization appears to be closer to
the value for metallic Co but to reach giant values at lower
concentrations. Unintended ferromagnetic signals like for the
substrates of course may also lead to ferromagnetic properties of
the thin TM doped ZnO film itself. Contamination with Fe from
handling, however, is rather unlikely since the face of the film
usually is not touched and ZnO itself is rather soft (table
\ref{tab:as_purch}). The other two unintended sources, i.e.
defects without involvement of the d-moments, as well as tiny
clusters, involving TM elements will be discussed in section 4 and
5.

\section{Role of the defects}
\subsection{Intrinsic defects in ZnO} Magnetic moments of p-like
electrons are by no means surprising. It is well known that
molecular oxygen develops paramagnetic moments which even couple
antiferromagnetically at low temperatures \cite{freiman}. Using
synchrotron radiation x-ray magnetic circular dichroism, recently
-electrons have been discovered to be the source for
ferromagnetism in proton irradiated carbon \cite{ohldag:187204}.
Since defects appear to be relevant for magnetic properties also
in ZnO, we would like to briefly review papers dealing with those
defects. One of the most instructive recent theoretical papers on
native point defects in ZnO was given by A. Janotti et al.
\cite{janotti:165202}. Here, the authors also give a critical
review of experimental results. Thus, we would like to summarize
the main results from their paper. The starting points are the six
possible point defects in the ZnO Wurtzite lattice, i.e. oxygen or
zinc vacancy, oxygen or zinc interstitial and oxygen or zinc
antisite. Although only six, the variety of defects is likely much
larger due to different possible charge states or pair as well as
complex formation. The authors made clear that two of them which
are often considered as candidates not only for n-type
conductivity but for mediation of ferromagnetic coupling, are
unlikely to be present in ZnO under equilibrium conditions. These
defects are the O vacancy (V$_O$) and the Zn interstitial
(Zn$_i$). Both V$_O$ and Zn$_i$ exhibit high formation energies in
n-type ZnO, even in extreme Zn-rich conditions, therefore, they
are unlikely to be formed. Moreover, V$_O$ is a deep rather than a
shallow donor with the (2+/0) transition level at $~$1 eV below
the bottom of the conduction band; therefore it cannot be
responsible for the often observed n-type conductivity in ZnO.
Note that oxygen vacancy in ZnO has been recently identified by
Vlasenko and Watkins \cite{vlasenko:125210} using optical detected
electron paramagnetic resonance (ODEPR). They emphasized that the
V$_O$ related EPR signals could be observed only after electron
irradiation. This is consistent with the fact that V$_O$ is not
present in ZnO under equilibrium, i.e., it has high formation
energy in n-type ZnO as predicted by Janotti et al.
\cite{janotti:165202}. Zinc interstitial (Zn$_i$) is a shallow
donor in ZnO. However, it exhibits a very low migration barrier,
thus, Zn$_i$ is mobile even below room temperature. Hence,
isolated Zn$_i$ is unstable and it is unlikely to be present in
detectable concentrations in n-type ZnO. The zinc vacancy
(V$_{Zn}$) has the lowest formation energy in n-type ZnO among all
intrinsic (native) point defects. V$_{Zn}$ is a deep acceptor and
acts as a compensating center in n-type ZnO. Interestingly,
Janotti et al. \cite{janotti:165202} proposed V$_{Zn}$ as one of
the causes of the often-observed green luminescence in ZnO. The
oxygen interstitial (O$_i$) also has a high energy. It is stable
as electrically inactive split interstitial O$_i$$^0$ (split) in
p-type ZnO, or as a deep acceptor at the octahedral site
O$_i$$^{1-}$(oct) in n-type ZnO. Zn antisite (Zn$_O$) is a shallow
donor, but it is unlikely to form in n-type ZnO due to its
exceedingly large energy of formation. Oxygen antisite (O$_{Zn}$)
has even higher formation energies and are also unlikely be
present in ZnO under equilibrium conditions. The authors also
provided temperatures, at which the defects are expected to become
mobile. The annealing temperatures are given in table
\ref{tab:point_defects}. The relevant point defects are predicted
to anneal anneal out at low or modest temperatures.

\begin{table}[b]\center
  \caption{Annealing temperatures of different native point defects in ZnO as pro-posed in Ref. \cite{janotti:165202}.}
  \begin{tabular}[htbp]{@{}ll@{}}
    \hline
    Defect & T$_{anneal}$ (K) \\
    \hline
    Zn$_i$$^{2+}$ & 219  \\
    V$_{Zn}$$^{2-}$ & 539   \\
    V$_O$$^{2+}$ &  655 \\
    V$_O$$^{0}$ &  909 \\
    O$_i$$^{0}$ (split) & 335 \\
    O$_i$$^{2-}$(oct) & 439 \\

    \hline
  \end{tabular}
  \label{tab:point_defects}
\end{table}

\begin{figure} \center
\includegraphics*[scale=0.35]{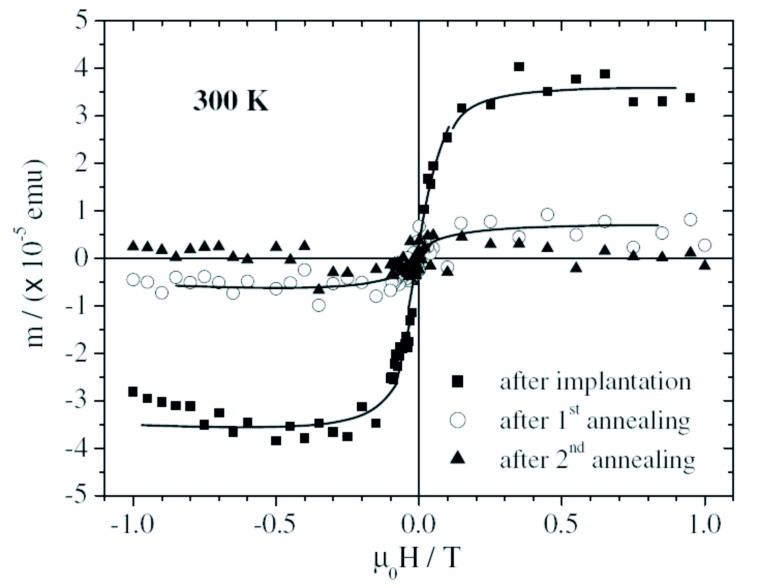}
\caption{%
Room temperature magnetization of the sample implanted with
1$\times10^{17}$ Ar cm$^{-2}$: after implantation (solid squares),
after annealing in air at 673 K for 6 h (open circles) and after a
second annealing in air at 773 K for 6 h (stars). Figure taken
from ref. \cite{borgesArZnO}}. \label{Fig4_2}
\end{figure}

Summarizing the paper we find that point defects are unlikely to
be associated to the often-observed n-type conductivity in ZnO.
Nevertheless, V$_O$ and Zn$_i$ can act as compensating centers in
p-type ZnO. The lowest energy defect in n-type ZnO is the zinc
vacancy which acts as deep acceptor, partially compensating the
n-type conductivity. On the other hand, point defects are formed
under non-equilibrium conditions. They become mobile and thus
might disappear at moderate temperatures or even room temperature.

An experimental study corresponding to those findings has been
performed by D. Sanyal et al. \cite{sanyal}. The group was
ball-milling ZnO powder and subsequently annealing it in air up to
873 K. Applying positron annihilation spectroscopy (PAS), they
found mainly Zn vacancies inside the as-milled powders. These
vacancies disappear upon annealing at 873 K while O-vacancies
start to appear at such high temperatures. Unlike metallic
clusters, defects very often appear, grow and disappear again due
to continuation or intensivation of certain treatment. Such
observation has been made by N. G. Kakazey et al. \cite{kakazey}
while grinding ZnO powder. Using EPR he identified different
paramagnetic defects dominated by Zn vacancy and O vacancy
formation. Note that due to the grinding process the signals of
the defects initially increase but decrease after a certain
grinding time. Finally we would like to focus shortly on the
creation of microscopic defects due to heavy destructive treatment
of the ZnO lattice. These defects are a result of agglomeration of
point defects and can lead to low dimensional structures within
ZnO. A very instructive example is given by the bombardment of
single crystalline ZnO with heavy ions \cite{kucheyev03}. After
bombardment of ZnO single crystals with Au ions at an energy of
300 keV and up to a fluence of 4${\times}10^{16}$ cm$^{-2}$ the
authors found
\begin{itemize}
    \item Oxygen loss close to the surface
    \item point defect clusters and
planar defects parallel to the basal plane of the ZnO wurtzite
structure
    \item Precipitate formation, likely Zn enriched regions
    \item No full amorphisation of the ZnO substrate reflecting dynamic
annealing
\end{itemize}

\subsection{Defects induced ferromagnetism}

\begin{figure} \center
\includegraphics*[scale=1.1]{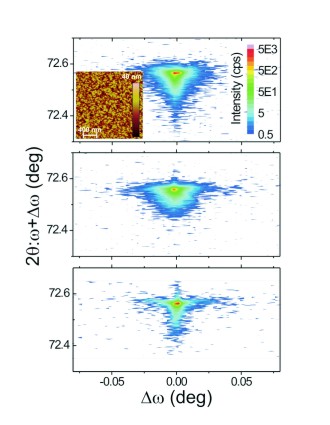}
\caption{%
HR-XRD at the ZnO(0004) reflection of (a) Fe(10\%):ZnO, (b)
XX:ZnO, and (c) a virgin sample from the same charge. (a) and (b)
show a similar diffuse background in the reciprocal space that is
much more pronounced than for a virgin sample (c). The inset in
(a) shows the surface of the corresponding AFM measurement
exhibiting self-organized dots created mainly due to the vacuum
pre-annealing. Figure modified from Ref. \cite{defects_zhou}.}.
\label{Fig4_3}
\end{figure}

The creation of ferromagnetic signals in pure, but defective ZnO
has been investigated recently by several groups, e.g., in laser
ablated ZnO thin films \cite{Hong07JPC}. In the paper of Borges et
al. \cite{borgesArZnO}, ferromagnetic properties have been
introduced in ZnO by means of Ar ion implantation. Although
spurious elements could be detected inside the crystals, the
authors could assign the ferromagnetic properties to defects
induced by the implantation. As in our papers published slightly
later (see below), the authors used reciprocal space mapping in
order to detect crystal lattice disorder. After implantation, the
crystals showed broadening of the (0004) diffraction peak
indicating enhanced strain or grain boundaries. Although being a
prerequisite for ferromagnetism, these microstructural defects are
not fully responsible for it since it disappears after annealing
at 773 K for a few hours (Fig. \ref{Fig4_2}).

Thus, the presence of a point defect appears to be the real source
of the observed ferromagnetism. The poisoning effect of
ferromagnetic signals originating from defects, however, is
largest if the ZnO is intentionally doped by TM ions in order to
create ferromagnetic DMS. A superb article on the field was
recently published by M. Gacic et al. \cite{gacic:205206} where Co
doped ZnO thin films were created by means of pulsed laser
deposition at a typical temperature of 873 K. The films exhibited
ferromagnetic properties at room temperature. Demonstrating the
large value of element specific electronic spectroscopy, the
authors proved, that the properties do not arise from the element
Co. Also no metallic Co clusters have been detected. In fact the
Co ions were paramagnetic. Thus, they associated the ferromagnetic
signal to defects. Consequently, also no indirect ferromagnetic
coupling between the Co ions is established by those defects. A
very similar experiment was performed by Xu et al.
\cite{xu:082508}. In this paper, Cu doped ZnO thin films have been
grown by PLD under nitrogen atmosphere. It was clearly proven,
that defects, rather than the Cu spins lead to ferromagnetic
properties. The origin of the ferromagnetism was assigned to Zn
vacancies rather than O vacancies. For our own investigation
investigation \cite{defects_zhou} we

\begin{figure} \center
\includegraphics*[scale=0.6]{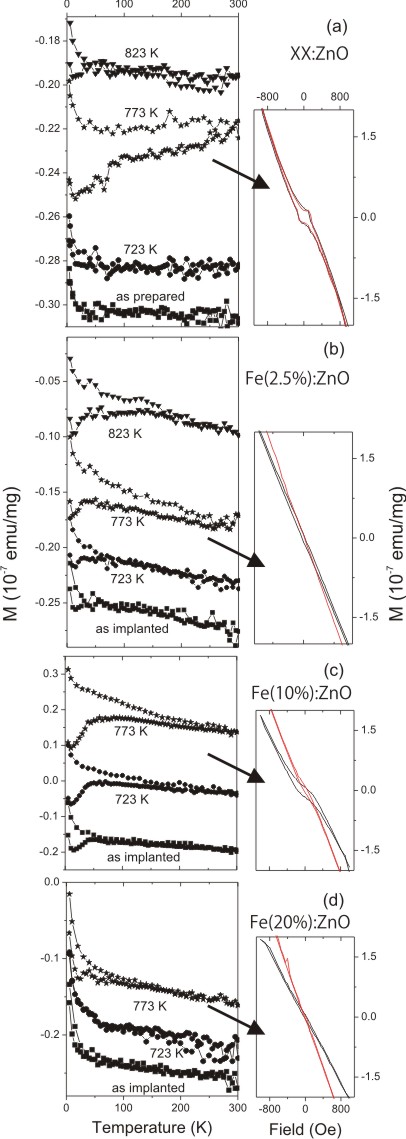}
\caption{%
ZFC/FC curves (left) and M vs. H dependence (right) for the
various crystals after selected preparation steps. The samples are
labelled according to Table I and the annealing temperatures are
indicated. The spectra - except the as-prepared/as-implanted - are
shifted in y-direction for better visibility. Black and red curves
(right) represent 5 K and 300 K measurement temperature,
respectively. Figure taken from Ref. \cite{defects_zhou}.}
\label{Fig4_5}
\end{figure}

\begin{itemize}
    \item Lowered the crystalline quality of the substrates intentionally by means of pre annealing them Zn face up in high vacuum at a temperature of 1073 K for 30 min. The temperature ramp was 60 K per minute for heating up and 2 K per minute for cooling down.
    \item Investigated the formation of
the delicate defect responsible for the ferromagnetic properties
by either implantation with Fe ions at different fluences or/and
post-annealing in vacuum at different temperatures.
\end{itemize}For the creation of the relevant defect we applied Fe$^+$ ion
implantation at a temperature of 253 K, an energy of 80 keV and
fluences of 0, 1, 4, and 8${\times}10^{16}$ cm$^{-2}$.
Corresponding to the maximum atomic concentration in the implanted
profile, we named the samples XX:ZnO, Fe(2.5\%):ZnO, Fe(10\%):ZnO,
and Fe(20\%):ZnO. The profiles were calculated using the TRIM
program \cite{trim}. The introduction of microscopic defects is
rather easy to be analyzed. Fig. \ref{Fig4_3} shows the RSM of a
pre-annealed crystal after implantation plus post-annealing and
the AFM picture of its surface. The surface roughness increased
from ~0.25 nm to 5 nm. RSM also reveals an increase of
strain/grain boundaries after such treatment \cite{defects_zhou}.
The profiles were calculated using the TRIM program \cite{trim}.
Fig. \ref{Fig4_5} instructively shows the behaviour of the
ferromagnetic properties and thus the defect due to implantation:
The splitting of the ZFC/FC curves recorded at 100 Oe initially
increases with increasing fluence but drops again at the largest
fluence. Moreover, we discovered that mild post-annealing in
vacuum of all of the samples leads to a similar effect like the
implantation. Exemplarily for XX:ZnO and Fe(2.5\%):ZnO [Fig.
\ref{Fig4_5} (a, b)] we observe an initial increase with
post-annealing temperature and finally a drop at higher
temperatures. This behavior fits the phenomenology of defect
induced ferromagnetism. Note that the drop of the thermomagnetic
irreversibility temperature T$_{irr}$, i.e. the bifurcation
temperature of ZFC and FC curves, is more drastic for XX:ZnO.
Moreover, we experienced a degradation of the ferromagnetic
properties for XX:ZnO after exposing it for a few days at ambient
conditions \cite{defects_zhou}.

In conclusion, defects can lead to unintended weak ferromagnetism
in oxides, \ie~in substrates, or deposited TM:ZnO films. The
ferromagnetic signals are not related to ordering of the TM 3d
moments. The major unintended ferromagnetic signals are, however,
secondary inclusions. We will mainly discuss our work in the
following section.

\section{Role of secondary phases}
\subsection{Introduction}The appearance of ferromagnetic secondary phases in TM
doped ZnO, a candidate for future spintronics devices, has been
investigated for the last 10 years. The reason is that such
clusters can lead to unwanted ferromagnetic signals that can be
confused with a "real" ferromagnetic DMS. On the other hand, the
clusters can influence necessary magneto-transport properties due
to the introduction of large amounts of metal-semiconductor
interfaces. One of the first papers on the characterization of
metallic clusters created from a popular TM dopant, Co, in ZnO was
given by Pearton et al \cite{norton03}. After implantation of 3-5
at.\% Co at a temperature of 623 K he found crystallographically
oriented hcp Co clusters to be present in the ZnO single crystals.
These clusters, although tiny, lead to room temperature
ferromagnetic properties of the whole sample. While these clusters
are rather "easy" to detect by means of x-ray diffraction, other
clusters require high resolution structural analysis (see below).
Especially clusters consisting only of a few atoms are nearly
impossible to be detected but might lead to pronounced magnetic
properties. Although we will not deal with such clusters here, it
is worth to mention two papers on the topic. The first,
theoretical article considers small (MnO)$_{x}$ clusters with
x$<$9 \cite{Nayak}. These clusters show exciting stability
dictated by their underlying magnetic configuration. The authors
found atomiclike magnetic moments ranging from 4 $\mu_B$ to 5
$\mu_B$ per MnO unit. Despite the antiferromagnetism in bulk MnO,
(MnO)$_x$ clusters with x=1-7, 9 show ferromagnetic coupling with
a localization of the moments at the Mn site. The same group
\cite{rao02} pointed out the relevance of N doped Mn clusters to
the ferromagnetic signal of Mn doped GaN. These clusters are
expected to exhibit moments of almost 5 $\mu_B$ per Mn. For TM
doped ZnO indeed small oxide inclusions have been found, namely
planar CuO nanophases of sizes ranging from 1 to 10 nm
\cite{sudakar:054423}. The authors prepared ZnO thin films doped
with $<$ 1 at.\% Cu exhibiting ferromagnetic properties by means
of radio-frequency sputtering. They related the ferromagnetic
signal to the presence of the embedded planar CuO nanophases. An
element specific prove of the relevance of the Cu 3d$^9$ moment
was not given. Nevertheless, the observed ferromagnetism was
assigned to uncompensated Cu 3d spins of the surface Cu ions.

\begin{figure} \center
\includegraphics*[scale=0.5]{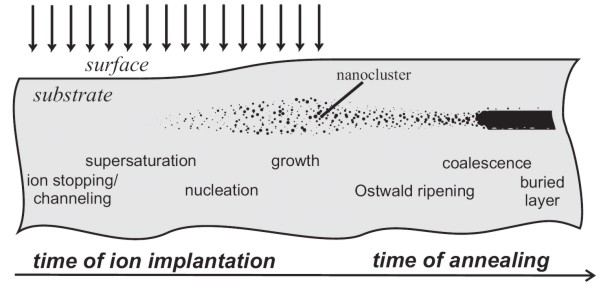}
\caption{%
Scheme of ion-beam synthesis of nanostructures \cite{Heinig03}.
High-fluence ion implantation into a surface layer produces
supersaturation of impurity atoms. Nano-clusters nucleate and grow
during ion implantation or, for impurity atoms immobile during ion
implantation, during subsequent annealing. The mean size of the
nanocluster as well as their spatial and size distribution changes
during Ostwald ripening. For very high ion doses, a buried layer
can form by coalescence.} \label{ostwald}
\end{figure}

Turning back to less exotic clusters we would like to review the
principles of the formation of secondary phases from a
supersaturated solid solution. It is usually created by
application of non-equilibrium preparation techniques like ion
implantation or low temperature thin film growth. Fig.
\ref{ostwald} shows a scheme of the formation of nanoclusters for
a supersaturated solid solution after high-fluence ion
implantation and subsequent annealing \cite{Heinig03}. Supposing
high enough mobility of the monomers and low enough energy of the
formation of a secondary phase, its nucleation can already happen
during implantation. Further implanting or post-implantation
annealing leads to the precipitation of clusters. The size
evolution of the clusters is described by Ostwald-ripening and
coalescence. For a very high ion implantation fluence, coalescence
of nano-clusters occurs, resulting in the formation of buried
layers (Fig. \ref{ostwald}).

Note that the phenomenon of spinodal decomposition has been
discussed recently \cite{sato05}. Spinodal decomposition describes
density fluctuations of the dopant driven by attractive chemical
pair interactions, e.g. of the dispersed Mn ions in GaMnAs or
GaMnN. These can lead to nanoscale regions of ferromagnetic DMS
due to enhancement of TM concentration. Spinodal decomposition
only occurs in a parameter window, where the mobility of the
dopant monomers is low and the energy of the formation of a second
phase cannot be reached, e.g. long-time exposure of the sample to
room temperature. The fluctuations in composition are relatively
small and thus different from nucleation which usually is
accompanied by large fluctuations of composition.

Actually, magnetic nanoclusters have been intentionally formed in
GaAs by ion beam synthesis \cite{wellmann97,shi95}. While these
clusters are "wanted", TM doped ZnO might show "unwanted" magnetic
second phases \cite{jin01}. In the following, we will discuss the
secondary phase formation in Fe, Co and Ni implanted ZnO, and
compare their differences in crystalline structures and in
annealing behaviors.


Before performing sample characterization, we first introduce the
magnetic behavior of a nanoparticle system. Such that we have some
clues to choose the analysis method.

For magnetic nanoparticles, a critical size may be reached, below
which the formation of magnetic domains becomes energetically
unfavorable. The critical diameter $d_c$ is given by

\begin{equation}\label{critical_diameter}
    d_c\approx18\frac{\sqrt{AK_{eff}}}{\mu_0M^2}
\end{equation} where $A$ is the exchange constant, $K_{eff}$ the effective anisotropy energy density, and $M$ the
saturation magnetization \cite{farle}. The critical diameter is 15
nm for Fe and 35 nm for Co \cite{farle}.

The magnetism of a single nanoparticle in a solid matrix follows
the N\'eel process \cite{respaud}. If the particle size is
sufficiently small, above a particular temperature (so-called
blocking temperature of T$_B$) thermal fluctuations dominate and
the particle can spontaneously switch its magnetization from one
easy axis to another. Such a system of superparamagnetic particles
does not show hysteresis in the M-H curves above T$_B$; therefore
the coercivity (H$_C$) and the remanence (M$_R$) are zero. Below
the blocking temperature, the particle magnetic moment is blocked
and its magnetization depends on its magnetic history.
Phenomenologically there are two characteristic features in the
temperature dependent magnetization of a nanoparticle system. One
is the irreversibility of the magnetization under a small applied
field (\eg~50 Oe) after zero field cooling and field cooling
(ZFC/FC) \cite{respaud}. The other is the drastic drop of the
coercivity and the remanence at a temperature close to or above
T$_B$ \cite{bean:S120}\cite{shinde04}.

\begin{figure} \center
\includegraphics[scale=0.75]{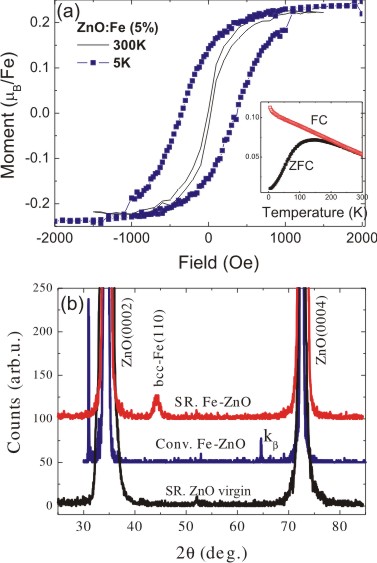}
\caption{\label{MH_Fe4E16}(a)Magnetization reversal recorded at 5
and 300 K using SQUID magnetometry for the sample implanted with
Fe at 623 K and the fluence of 4$\times$10$^{16}$ cm$^{-2}$. Inset
shows the ZFC/FC curves with an applied field of 50 Oe for the
same sample. (b) Conventional (Conv.) and SR-XRD pattern
($2\theta$-$\theta$ scan) for the Fe implanted ZnO. A virgin
sample is shown for comparison. Adapted from Ref.
\cite{zhouFe}.}\label{fig:Fe4E16}
\end{figure}

Obviously one question is whether one can judge the origin of
ferromagnetic properties from the temperature dependence of ZFC/FC
magnetization. At least, a system of DMS nanoparticles,
\eg~nanostructured Mn-doped InP \cite{poddar:062506}, behaves
exactly the same as normal magnetic nanoparticles. Moreover,
Roshko \etal~used the Preisach model in order to calculate the
ZFC/FC magnetization for a conventional ferromagnet
\cite{Roshko00}. The thermal fluctuation energy in a ferromagnet
is very small so that blocking and activation only occur very
close to Curie temperature (T$_C$). They found that ZFC/FC curves
have similar shape as that of a superparamagnetic nanoparticle
system, but with a maximum in ZFC curve just below T$_C$.
Additionally some frustrated systems, \eg~a spin glass, also show
slow dynamical behavior, which results in the irreversibility of
the magnetization after zero field cooling and field cooling
\cite{takano:8197}. Therefore, it is impossible to unambiguously
judge the magnetic origin, \ie~secondary phase,
spinodal-decomposition, nano-DMS, and spin-glass, by ZFC/FC
magnetization measurement alone. For this purpose one has to
correlate the structural and magnetic properties. Since the
majority of the groups are using XRD for structural
characterization, we will focus on that technique discussing our
results in more detail. We used both synchrotron radiation (SR)
XRD and conventional XRD.

Structural analysis was achieved both by synchrotron radiation
x-ray diffraction (SR-XRD) and conventional XRD. SR-XRD was
performed at the Rossendorf beamline (BM20) at the ESRF with an
x-ray wavelength of 0.154 nm. Conventional XRD was performed with
a Siemens D5005 equipped with a Cu-target source. In XRD
measurement, we use 2$\theta$-$\theta$ scans to identify
crystalline precipitates, and pole figures (azimuthal $\phi$-scan)
for determining their crystallographical orientation.

\subsection{Metallic secondary phases (Fe, Co, Ni)}

\begin{figure}\center
\includegraphics[scale=0.4]{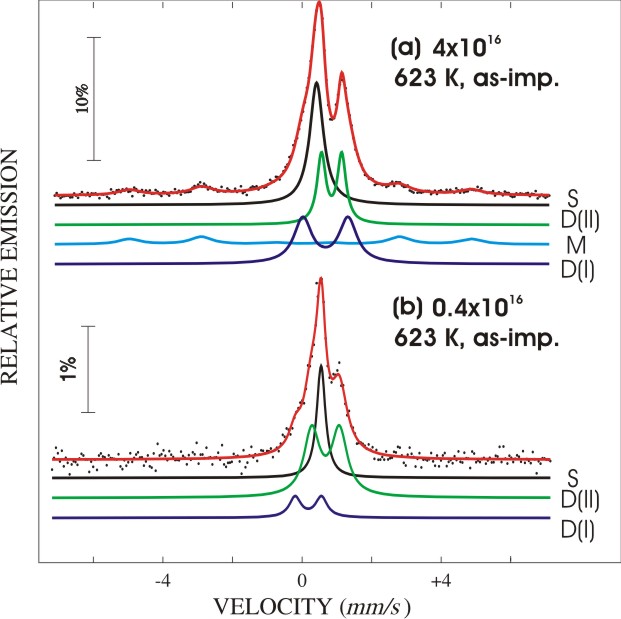}
\caption{Room temperature CEMS of ZnO bulk crystals implanted with
$^{57}$Fe different fluences at a temperature of 623 K. The
notations for the fitting lines are given as S (singlet), D
(doublet) and M (sextet). Adapted from Ref.
\cite{zhouFe}.}\label{fig:CEMS_fluence_ZnO}
\end{figure}

\subsubsection{Fe implanted ZnO} We pick out the ZnO single crystals implanted with Fe as an example
to show the possible misinterpretation of the observed
ferromagnetism in transition metal implanted ZnO. Fig.
\ref{fig:Fe4E16}(a) shows the magnetization measurement on the
sample implanted with Fe, with the field along the sample surface.
The implantation temperature is 623 K and the Fe fluence is
4$\times$10$^{16}$ cm$^{-2}$. This fluence corresponds to 5\%
maximum atomic concentration of the Gauss-like shaped Co$^+$
implantation profile \cite{trim}. Note that for ion implantation
the concentration usually is given with respect to the total
atomic density and not as x from Zn$_{1-x}$TM$_x$O. At both 5 K
and 300 K, the sample shows ferromagnetism. However with
increasing temperature, its coercivity and remanence are decreased
drastically: from 360 Oe at 5 K to 10 Oe at 300 K, and 0.14
$\mu_B$/Fe to 0.01 $\mu_B$/Fe, and this is a strong indication of
superparamagnetism, which has been confirmed by the measurement of
ZFC/FC magnetization. The inset of Fig. \ref{fig:Fe4E16}(a)  shows
ZFC/FC curves with an applied field of 50 Oe. A distinct
difference in ZFC/FC curves was observed. ZFC curves show a
gradual increase (deblocking) at low temperatures, and reach a
broad peak with a maximum, while FC curves continue to increase
with decreasing temperature. The broad peak in the ZFC curves is
due to the size distribution of Fe NCs.

\begin{figure} \center
\includegraphics[scale=0.7]{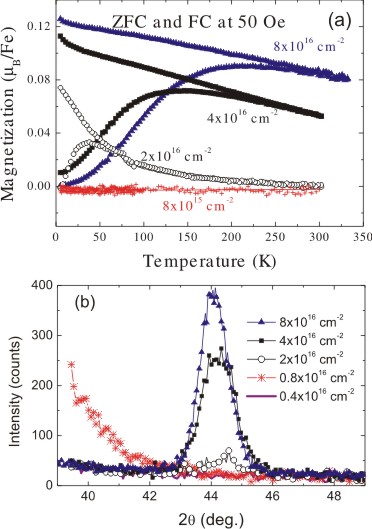}
\caption{(a) Magnetization curves with an applied field of 50 Oe
after ZFC/FC for the Fe implanted ZnO. The lower branch is ZFC
curve while the upper branch is FC curve. With increasing fluence,
the Fe NCs are growing in size, resulting in a higher blocking
temperature. (b) Hysteresis loops measured at 5 K for Fe implanted
ZnO with different fluence. Adapted from Ref.
\cite{zhouFe}.}\label{fig:SQUID_FeZnO_fluence}
\end{figure}

Of course in order to confirm the existence of nanoscale metallic
Fe precipitates one needs structural support. Fig.
\ref{fig:Fe4E16}(b) shows symmetric 2$\theta$/$\theta$ scans for
the sample performed at conventional XRD and SR-XRD. Obviously no
secondary phases could be detected by conventional XRD, where the
sharp peaks, at 2$\theta$$\sim$34.4\degr~and
2$\theta$$\sim$72.6\degr, are from bulk ZnO. In contrast to
conventional XRD, the much higher x-ray intensity in SR-XRD allows
one to detect also small amounts of very tiny nanoparticles.  At
2$\theta$$\sim$44.5\degr, a rather broad and low intensity peak
originating from alpha-Fe(110) with a theoretical Bragg angle of
2$\theta$=44.66\degr~occurs. The nanoparticle size is estimated to
be around 8 nm using the Scherrer formula \cite{scherrer}. Apart
from alpha-Fe, no other Fe-oxide (Fe$_2$O$_3$, Fe$_3$O$_4$, and
ZnFe$_2$O$_4$) particles are detected. By combining the magnetic
and structural measurements, it is reasonable to conclude that
metallic $\alpha$-Fe nanoparticles have formed upon implantation
at 623 K with the fluence of 4$\times$10$^{16}$ cm$^{-2}$, and
they are responsible for the ferromagnetism. The saturation moment
at 5 K is around 0.24 $\mu_B$/Fe. By comparing with the bulk Fe
with a saturation magnetization of around 2.2 $\mu_B$/Fe, around
11\% of Fe in this sample is in metallic state. This is further
confirmed by CEMS measurement.

Fig. \ref{fig:CEMS_fluence_ZnO}(a) show CEMS spectrum for Fe
implanted ZnO at 623 K with a fluence of 4$\times$10$^{16}$
cm$^{-2}$. The majority of Fe are ionic states Fe$^{3+}$ (singlet
S) and Fe$^{2+}$ (doublet D(I) and D(II)), while a considerable
fraction of a sextet associated to $\alpha$-Fe is present (sextet
M). The amount of metallic Fe obtained from CEMS simulation is
12.5\%. It is in quite good agreement with the results by
magnetization measurement. On the other hand, these Fe$^{2+}$ and
Fe$^{3+}$ ions could be dispersed inside ZnO matrix.

Now we would like to find out, if lowering of the fluence results
in the avoidance of metallic secondary phases. Fig.
\ref{fig:SQUID_FeZnO_fluence}(a) shows the ZFC/FC magnetization
curves in a 50 Oe field for different fluences of Fe implanted
ZnO. The FC curves for low fluences of 1$\times$10$^{15}$ (not
shown to avoid overlap) and 8$\times$10$^{15}$ cm$^{-2}$
completely overlap with the corresponding ZFC curves at zero
level. No superparamagnetic particles are present in the two
samples. For larger fluences (above 2$\times$10$^{16}$ cm$^{-2}$),
the irreversibility in ZFC/FC curves was observed.

Correspondingly SR-XRD reveals the same trend for metallic Fe
formation. Fig. \ref{fig:SQUID_FeZnO_fluence}(b) shows the SR-XRD
pattern (focused on Fe(110) peak) as a function of fluence. At a
low fluence (1$\times$10$^{15}$ to 8$\times$10$^{15}$ cm$^{-2}$),
no crystalline Fe NCs could be detected, while above a fluence of
2$\times$10$^{16}$ cm$^{-2}$), an Fe(110) peak appears and
increases with fluence. The inset shows a wide range scan for the
high fluence sample (4$\times$10$^{16}$ cm$^{-2}$). The full width
at half maximum (FWHM) of the Fe(110) peak decreases with fluence,
indicating a growth of the average diameter of these NCs,
according to the Scherrer formula \cite{scherrer}. Note, that not
only the presence of tiny metallic TM clusters is relevant for
magnetic properties but also the crystallographic orientation with
respect to the host matrix. However, no texture behavior is found
even for the highest fluence sample in pole figure measurements on
Fe(110) and Fe(200) (not shown). This could be due to the
difference in the crystalline symmetry of hexagonal ZnO (six fold
symmetry) and bcc-Fe (four fold symmetry). For a bcc-crystal, one
cannot find a six-fold symmetry viewed from any direction. In
contrast, hcp-Co(0001) and fcc-Ni(111) NCs, which are six-fold
symmetric, are found to be crystallographically oriented inside
ZnO matrix. This will be shown in the next section.

As a cross-check to confirm no metallic Fe in the samples with low
Fe fluence, Fig. \ref{fig:CEMS_fluence_ZnO}(b) shows the CEMS
spectrum of Fe implanted ZnO at 623 K with a fluence of
0.4$\times$10$^{16}$. The singlet S and doublet D(I) are
attributed to Fe$^{3+}$, while the doublet D(II) is from
Fe$^{2+}$. No Fe$^0$ state could be detected.

With post-implantation annealing, one expects that the metallic Fe
nanoparticles grow driven by Ostwald ripening. According to
magnetization measurement, we found that the annealing at 823 K
results in the growth of $\alpha$-Fe nanoparticles. During
annealing at 1073 K the majority of the metallic Fe is oxidized;
after a long term annealing at 1073 K, crystallographically
oriented ZnFe$_2$O$_4$ NCs form, which will be discussed in the
section \ref{section:ZnFeO}.

\subsubsection{Co and Ni implanted ZnO} In this section, the formation of Co and Ni
nanocrystals inside ZnO upon implantation will be discussed. Co or
Ni ions were implanted into ZnO at 623 K with the fluence from
$0.8\times10^{16}$ cm$^{-2}$ to $8\times10^{16}$ cm$^{-2}$. The
maximum atomic concentration thus ranges from $\sim$1\% to
$\sim$10\%. Already in the as-implanted samples, Co or Ni NCs have
formed, and they exhibit superparamagnetic properties.

\begin{figure} \center
\includegraphics[scale=0.7]{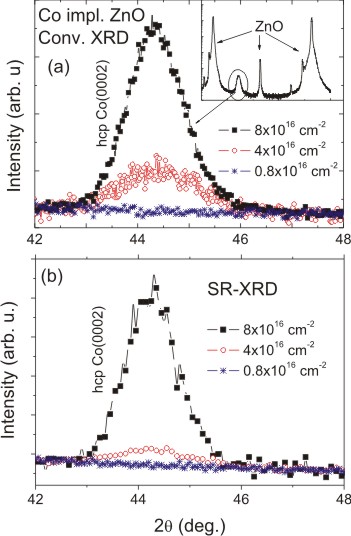}
\caption{2$\theta$-$\theta$ scan revealing the formation of
metallic Co precipitates in Co implanted ZnO (a) conventional XRD,
and (b) synchrotron XRD. (The fluence for Co ions is indicated).
Adapted from Ref. \cite{zhou07CoNi}. }\label{fig:XRD_CoZnO}
\end{figure}

Both conventional and synchrotron XRD techniques were employed to
check the formation of secondary phases in Co or Ni implanted ZnO.
Obviously conventional XRD already can detect the formation of
metallic Co nanocrystals as shown in Fig. \ref{fig:XRD_CoZnO}(a).
At a low fluence (0.8$\times$10$^{16}$ cm$^{-2}$), no crystalline
Co nanocrystals could be detected. At large fluences starting from
4$\times$10$^{16}$ cm$^{-2}$ the hcp-Co(0002) peak appears (Using
approaches presented in Ref. \cite{zhou07CoNi} we are able to
identify the Co phase to be hcp and not fcc). SR-XRD reveals the
same fluence dependence of Co nanocrystals [Fig.
\ref{fig:XRD_CoZnO}(b)]. XRD reveals similar results for Ni
implanted ZnO (not shown), \ie~Ni NCs start to form at the Ni
fluence of 4$\times$10$^{16}$ cm$^{-2}$.

\begin{figure} \center
\includegraphics[scale=0.55]{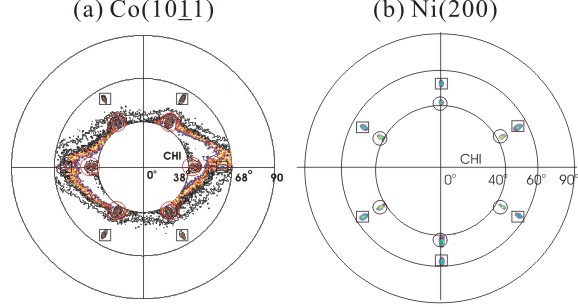}
\caption{XRD Pole figure revealing the crystallographical
orientation relationship between Co/Ni NCs and ZnO matrix, (a)
Co(10$\underline{1}$1) (in square) together with the tail of
ZnO(10$\underline{1}$2) (in circle); (b) Ni(200) (in square)
together with the tail of ZnO(10$\underline{1}$2) (in circle).
From Ref. \cite{zhou07CoNi}.}\label{fig:CoNiZnO_polefigure}
\end{figure}

Fig. \ref{fig:CoNiZnO_polefigure} (a) and (b) show the pole figure
of hcp-Co(10$\underline{1}$1) and fcc-Ni(200), respectively. The
radial coordinate is the angle ($\chi$) by which the surface is
tilted out of the diffraction plane. The azimuthal coordinate
($\phi$) is the angle of rotation about the surface normal. The
pole figure shows poles of hcp-Co(10$\underline{1}$1) at
$\chi$$\sim$61.9\degr, and Ni(200) at $\chi$$\sim$54.8\degr,
respectively. Both exhibit a sixfold symmetry. Since
ZnO(10$\underline{1}$2) and hcp-Co(10$\underline{1}$1) have
similar Bragg angle, the poles of ZnO(10$\underline{1}$2) also
show up at $\chi$$\sim$42.8\degr~with much more intensities. The
results are consistent with the theoretical Co(10$\underline{1}$1)
pole figure viewed along [0001], and Ni(200) pole figure viewed
along [111] direction, respectively. Therefore, we can conclude
that these Co and Ni NCs are crystallographically oriented with
respect to the ZnO matrix. The in-plane orientation relationship
is hcp-Co[10$\underline{1}$0]$\parallel$ZnO[10$\underline{1}$0],
and Ni[112]$\parallel$ZnO[10$\underline{1}$0], respectively. Due
to the hexagonal structure of Co and sixfold symmetry of Ni viewed
along [111] direction, it is not difficult to understand their
crystallographical orientation onto hexagonal-ZnO.

\begin{figure} \center
\includegraphics[scale=0.7]{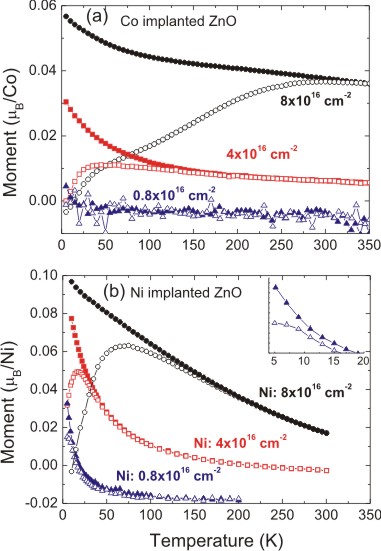}
\caption{Magnetization curves at 50 Oe after ZFC/FC (a) Co
implanted ZnO; (b) Ni implanted ZnO, Inset shows a zoom of the low
temperature part of the ZFC/FC for the fluence of
0.8$\times$10$^{16}$ cm$^{-2}$, which reveals the similar
behaviour as higher fluence sample, but with a T$_B$ $\leq$ 5 K.
For both series of samples, the Co/Ni NCs are growing in size with
increasing fluence, resulting in a higher blocking temperature;
Adapted from Ref. \cite{zhou07CoNi}.}\label{fig:ZFCFC_CoNiZnO}
\end{figure}

Correspondingly, magnetization measurements reveal similar fluence
dependence of the formation of metallic Co or Ni nanocrystals.
Fig. \ref{fig:ZFCFC_CoNiZnO}(a) shows the temperature dependent
magnetization curves after ZFC/FC with H$=50$ Oe. Knowing the
formation of hcp-Co from XRD, it is reasonable to assume that
hcp-Co NCs are responsible for the magnetic behavior. For bulk
hcp-Co crystals, the magnetic moment is 1.7 $\mu_{B}$/Co at 0 K.
Assuming the same value for Co NCs and using the saturation
magnetization measured at 5 K, around 17\% and 26\% of implanted
Co ions are in the metallic state for the fluence of 4$\times$ and
8$\times$10$^{16}$ cm$^{-2}$, respectively. Similar results are
observed for Ni implanted ZnO. Fig. \ref{fig:ZFCFC_CoNiZnO}(b)
shows the ZFC/FC magnetization curves for Ni implanted ZnO with
different fluences. Comparing with Co, Ni has a much lower
anisotropy energy density. For similar sizes of Ni NCs, the
blocking temperature is therefore much lower than that of Co.

\begin{figure} \center
\includegraphics[scale=0.4]{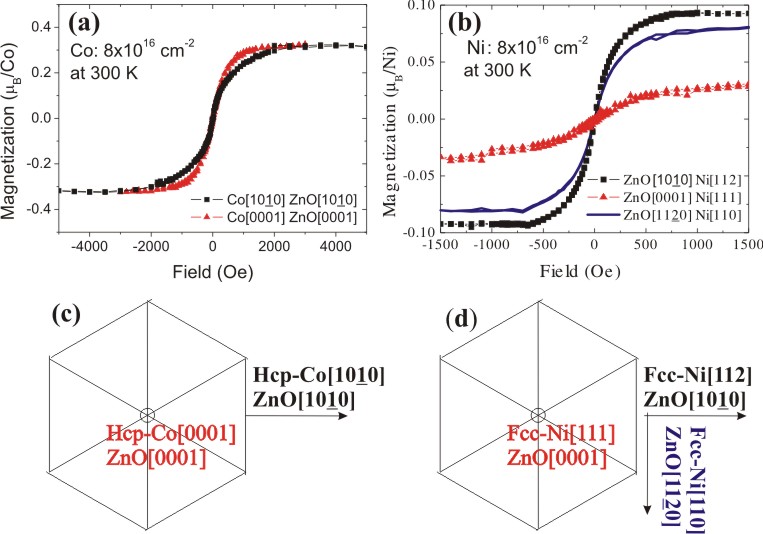}
\caption{Hysteresis loops measured with the field along ZnO[0001]
(out-of-plane) and [10$\underline{1}$0] (in-plane) for Co/Ni
implanted ZnO with the fluence of 8$\times$10$^{16}$ cm$^{-2}$
measured at 300 K, (a) Co implanted ZnO; and (b) Ni implanted ZnO.
(c) and (d) show the schematic geometry for magnetization
measurements. Obviously Co[0001] is the easy axis, the same as a
bulk hcp-Co crystal. However for Ni nanocrystals, the easy axis is
Ni[112] and the hard axis is Ni[111], which is just the opposite
for bulk Ni. Adapted from Ref.
\cite{zhou07CoNi}.}\label{fig:CoNiZnO_aniso_asimp}
\end{figure}

\begin{figure} \center
\includegraphics[scale=0.7]{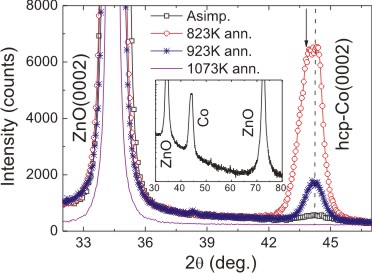}
\caption{XRD 2$\theta$-$\theta$ scans for Co implanted ZnO
crystals with different annealing temperature. The Co fluence is
4$\times$10$^{16}$ cm$^{-2}$. The wide range XRD pattern for one
of the samples (inset) reveals that no other crystalline phase
(e.g. CoO) could be detected. The arrow points the peak shoulder
coming from fcc-Co(111) diffraction in the sample of 823 K ann.
From Ref. \cite{zhou07CoNi}.}\label{fig:XRD_CoNiZnO_ann}
\end{figure}

Aiming at special features of magnetic anisotropy, Fig.
\ref{fig:CoNiZnO_aniso_asimp}(a) shows the comparison of the
magnetization along ZnO[10$\underline{1}$0] and [0001] at 300 K
for Co and Ni, respectively. Fig. \ref{fig:CoNiZnO_aniso_asimp}(c)
shows the orientation relationship between hcp-Co and ZnO, and the
measurement geometry. Obviously Co[0001] is the easy axis, the
same as a bulk hcp-Co crystal. The intersection of both curves
gives an effective anisotropy field of 3000 Oe. Fig.
\ref{fig:CoNiZnO_aniso_asimp}(b) shows the same measurement of Ni
implanted ZnO, while (d) shows the orientation relationship
between fcc-Ni and ZnO, and the measurement geometry. In contrast
to bulk Ni where [111] is the easy axis, here the easy axis is
Ni[112] and the hard axis is Ni[111]. Moreover, as shown in Fig.
\ref{fig:CoNiZnO_aniso_asimp}(b), another in-plane direction
Ni[110] is also an easy axis. Within the applied field, the
magnetization curve along the hard axis does not intersect with
that along the easy axis. The effective anisotropy field is much
larger than 1500 Oe. That means that there are other contributions
to the anisotropy dominating over the crystalline magnetic
anisotropy. By XRD measurement (Fig. \ref{fig:XRD_CoNiZnO_ann}),
we can evaluate the lattice constant of Co or Ni nanocrystals.
Actually we find both of them are tensilely strained in the
perpendicular direction. This lattice strain results in the
magnetoelastic energy, which could over-dominated the
magnetocrystalline anisotropy energy. Comparing with hcp-Co, Ni
has a much smaller (two orders of magnitude) magnetocrystalline
anisotropy constant. Therefore the magnetoelastic anisotropy
energy dominates the total anisotropy energy, resulting in the
in-plane anisotropy. On the other hand, the hcp-Co NCs in the
as-implanted sample still keep the bulk like anisotropy behavior.

In order to further investigate the Co implanted ZnO,
post-annealing was also performed in high vacuum at temperatures
ranging from 823 K to 1073 K for 15 min for the sample with Co
fluence of 4$\times$10$^{16}$ cm$^{-2}$. Fig.
\ref{fig:XRD_CoNiZnO_ann} shows the development of Co NCs upon
thermal annealing. The peak area and crystallite size calculated
using the Scherrer formula \cite{scherrer} are compared in table
I. A broad scan (the inset of Figure \ref{fig:XRD_CoNiZnO_ann})
reveals only one peak from Co besides the ZnO peaks. Using an XRD
$\phi$-scan (see Ref. \cite{zhou07CoNi}), we find only hcp-Co in
the as-implanted sample and the sample annealed at 923 K, while
both fcc- and hcp-Co are present in the sample annealed at 823 K.
Note that the peak area of Co in Figure \ref{fig:XRD_CoNiZnO_ann},
which is an approximate measure of the amount of Co NCs, increases
drastically after 823 K annealing, while decreases after 923 K
annealing. It is reasonable to attribute this change to the
formation and disappearance of fcc-Co. The fcc-Co is probably
oxidized to the amorphous CoO after 923 K annealing, while finally
all Co NCs are oxidized to an amorphous state after annealing at
1073 K.

\begin{figure} \center
\includegraphics[scale=0.7]{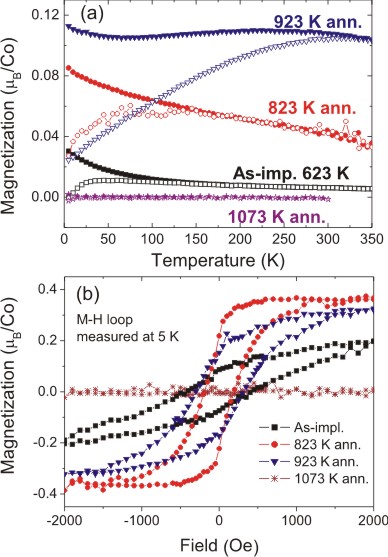}
\caption{(a) ZFC/FC magnetization curves at 50 Oe for the samples
after Co implantation and annealing at different temperatures.
Solid symbols are FC curves, while open symbols are ZFC curves.
(b) M-H curves for the same set of samples. The Co fluence is
4$\times$10$^{16}$ cm$^{-2}$, and the annealing temperatures are
indicated. Adapted from Ref.
\cite{zhou07CoNi}.}\label{fig:SQUID_CoZnO_ann}
\end{figure}

The structural phase transformation of Co NCs results in
corresponding magnetic properties as revealed by SQUID. Fig.
\ref{fig:SQUID_CoZnO_ann}(a) and (b) show the ZFC/FC magnetization
curves and magnetization loops, respectively, for all samples
annealed at different temperatures. No significant magnetization
response is detected for the sample annealed at 1073 K. The
saturation magnetization is increased after annealing at 823 K,
while decreased again after annealing at 923 K. This reflects the
formation and oxidation of fcc-Co as found by XRD (see Fig.
\ref{fig:XRD_CoNiZnO_ann}).

Similar annealing results were found for Ni implanted sample (not
shown). The mild temperature annealing (823 K) only slightly
increases the grain size of Ni. The annealing at 923 K drastically
decreases the peak area, while the grain size also decreases. The
Ni NCs were completely oxidized to amorphous state after 1073 K
annealing, and thus are non-detectably by XRD.

\subsection{Oxide secondary phases}\label{section:ZnFeO}
Oxide precipitates of TM are usually expected upon preparation at
large oxygen partial pressure. Consequently, a ferromagnetic,
metastable, oxygen-vacancy-stabilized
Mn$_{2-x}$Zn$_{x}$O$_{3-\delta}$ phase have been identified in Mn
doped ZnO created by solid state reaction \cite{kund04}. Possible
simple secondary oxide phases of transition metals occurring in TM
doped ZnO are mentioned in Ref. \cite{jin01}. As a bulk material,
some of them are antiferromagnetic, like MnO, NiO, and CoO. Others
are ferromagnetic or ferrimagnetic, like CrO$_2$, Fe$_3$O$_4$, and
$\gamma$-Fe$_2$O$_3$. Later on Zhou \etal~using Raman spectroscopy
found the existence of antiferromagnetic cobalt oxides like CoO
and Co$_3$O$_4$ already in intermediately doped ZnO
\cite{zhouhj06}. Nayak \etal~applied first-principles calculation
and Monte Carlo simulation of a classical Heisenberg model for Co
doped ZnO and found the formation of cobalt oxide system for
larger Co concentrations \cite{nayak08}. Taking into account
nanoscale precipitates, the situation becomes more complicated.
E.g. non-compensated spins at the surfaces of antiferromagnetic
clusters might provide substantial ferromagnetic signal
\cite{dietl:155312}.

Another source for ferromagnetic signals are inverted spinels
involving both Zn and O elements. For example, ferrimagnetic
spinel (Zn,Mn)Mn$_{2}$O$_{4}$ has been found in Mn doped ZnO
\cite{zheng04}. The sample has a very large coercivity of 5500 Oe
at 5.5 K and a Curie temperature of 43 K. Exchange bias is clearly
observed below 22 K, which is attributed to the exchange
interaction between ferrimagnetic (Zn,Mn)Mn$_{2}$O$_{4}$ and
spin-glass-like (or antiferromagnetic) phase in manganese oxides.
Another example is ZnCo$_2$O$_4$. By Raman spectroscopy
ZnCo$_2$O$_4$ was identified in Co doped both ZnO nanostructures
\cite{wang:031908} and thin films \cite{samanta:245213}. None of
the two papers attributed the observed ferromagnetism to spinel
ZnCo$_2$O$_4$, which, however, does show ferromagnetic behavior
\cite{kim:7387}. Concerning our own investigation, we show the
synthesis of Zn-ferrites by Fe implantation into ZnO and
post-annealing. By this we would like to point out that oxides NC
can account for a ferromagnetic signal, although a bulk material
can be antiferromagnetic.

\subsubsection{ZnFe$_2$O$_4$ nanocrystals}

\begin{figure}\center
\includegraphics[scale=0.7]{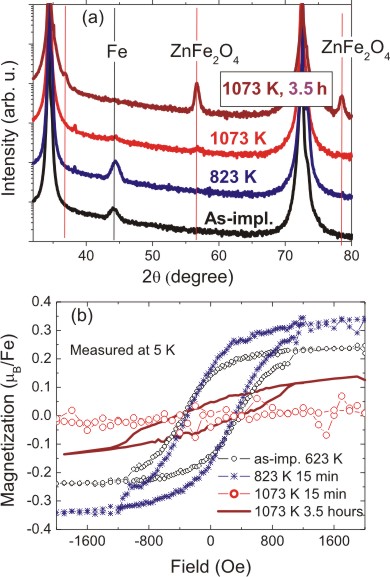}
\caption{(a) SR-XRD patterns (2$\theta$-$\theta$ scans) of Fe
implanted ZnO reveal the second phase development (from
$\alpha$-Fe to ZnFe$_2$O$_4$) upon annealing for 15 min (except
the 3.5 hours indicated) at different temperatures. (b)
Magnetization versus field reversal revealing the magnetism
evolution upon annealing. Adapted from Ref. \cite{zhou07JPD}.}
\label{fig:Fig1ZnFeO}
\end{figure}

Figure \ref{fig:Fig1ZnFeO}(a) shows the SR-XRD patterns for the
as-implanted and annealed samples. After annealing at 823 K for 15
min, larger and more Fe nanoparticles are formed as compared to
the as-implanted state. This is reflected by an increase and the
sharpening of the corresponding peak at 44.4\degr~in the
2$\theta$-$\theta$ scan. After annealing at 1073 K for 15 min, the
Fe(110) peak almost disappeared and the sample already shows an
indication for the presence of ZnFe$_2$O$_4$. After 3.5 hours
annealing at 1073 K, crystalline and oriented ZnFe$_2$O$_4$
particles are clearly identified.

\begin{figure}\center
\includegraphics[scale=0.4]{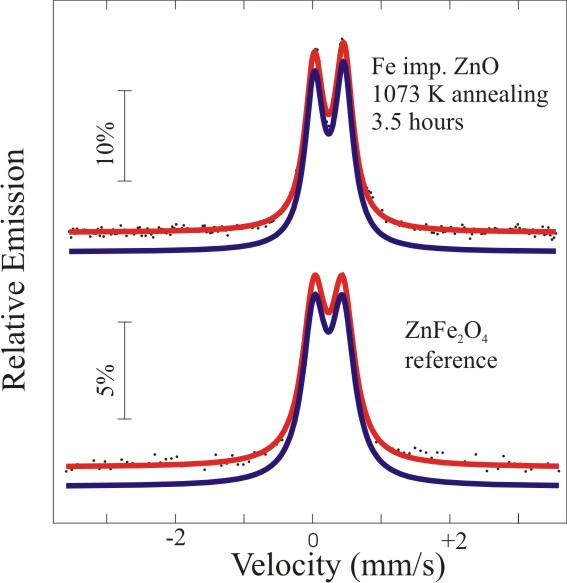}
\caption{Room-temperature CEMS of Fe implanted ZnO after annealing
at 1073 K for 3.5 hours. A reference sample of powder
ZnFe$_2$O$_4$ is also shown for comparison.}\label{fig:Fig2ZnFeO}
\end{figure}

The formation of Zn-ferrites is further confirmed by CEMS (Figure
\ref{fig:Fig2ZnFeO}). After 3.5 hours annealing, the CEMS pattern
exhibits only one quadrupole split line typical for ZnFe$_2$O$_4$
at room temperature \cite{chinnasamy00,hamdeh:1851}.


Figure \ref{fig:Fig1ZnFeO}(b) shows the magnetization versus field
reversal (M-H) at 5 K. Magnetic hysteretic loops are observed for
the as-implanted sample, which contains  $\alpha$-Fe
nanoparticles. After annealing at 823 K for 15 min, the
ferromagnetism is enhanced, \ie~Fe nanoparticles are growing in
size and amount. However, after annealing at 1073 K for 15 min, no
hysteresis loop is observed. Probably the majority of Fe particles
were oxidized to some amorphous nonmagnetic compound, as in the
case for Ni and Co, also at 1073 K annealing. The magnetism
evolution is in a good agreement with the XRD measurement. After
annealing at 1073 K for 3.5 hours, the hysteretic behavior is
observed again, which - according to the SR-XRD - cannot come from
Fe nanoparticles. It results from partially inverted
ZnFe$_2$O$_4$, which is ferrimagnetic
\cite{chinnasamy00,hamdeh:1851,zhou:3167}, in contrast to bulk
ZnFe$_2$O$_4$. The explanation for such behavior is additional
occupation of tetrahedral A sites by Fe and octahedral B sites by
Zn (as shown in Eq. \ref{normal} and \ref{partial}). The
intra-site interaction of the magnetic moment of the cations on
the B sites is much weaker than the A–B inter-site one. The
Fe$^{3+}$ ions between A and B sites couple antiferromagnetically,
which leaves some net spins from the non A-B paired Fe$^{3+}$
ions.

Normal spinel: \begin{equation}\label{normal}
(Zn^{2+}){\cdot}(\overrightarrow{Fe^{3+}}\cdot\overleftarrow{Fe^{3+}}){\cdot}O_{4}
\end{equation}
Inverted spinel:
\begin{equation}\label{partial}
(\overrightarrow{Fe^{3+}}{\cdot}Zn^{2+})\cdot(\overleftarrow{Fe^{3+}}{\cdot}Zn^{2+}\cdot\overrightarrow{Fe^{3+}}){\cdot}O_{4}
\end{equation}

\begin{figure}\center
\includegraphics[scale=0.45]{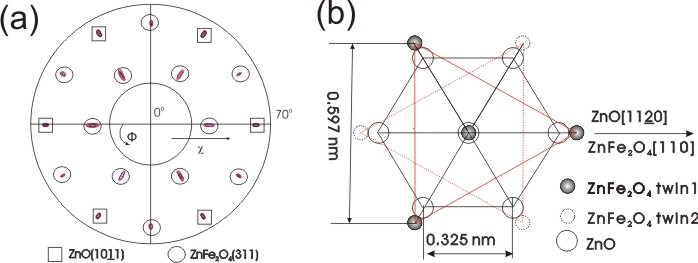}
\caption{(a) Pole figure of ZnFe$_2$O$_4$(311) reveals the
crystallographical orientation of ZnFe$_2$O$_4$ and its twin
crystallites. (b) A schematics for the crystallographical
orientation of ZnFe$_2$O$_4$ onto ZnO. From Ref. \cite{zhou07JPD}.
From Ref. \cite{zhou07JPD}.}\label{fig:Fig3ZnFeO}
\end{figure}

In Fig. \ref{fig:Fig1ZnFeO}(a), the XRD pattern for the sample
after 3.5 hours annealing at 1073 K shows only three peaks of
ZnFe$_2$O$_4$(222) (333) and (444), which means that the
crystallites of ZnFe$_2$O$_4$ are not randomly, but highly
oriented. The surface orientation is
ZnFe$_2$O$_4$(111)$\parallel$ZnO(0001). The crystallographical
orientation of ZnFe$_2$O$_4$ was revealed by the XRD pole figure.
Fig. \ref{fig:Fig3ZnFeO}(a) shows the pole figure for
ZnFe$_2$O$_4$(311). Poles of ZnFe$_2$O$_4$(311) at
$\chi$$\sim$29.5\degr and 58.5\degr, respectively, with sixfold
symmetry, are visible. Since ZnO(1011) (2$\theta$=36.25\degr) has
a close Bragg angle with ZnFe$_2$O$_4$(311)
(2$\theta$=35.27\degr), the poles of ZnO(10$\underline{1}$1) also
show up at $\chi$$\sim$61.6\degr~with much more intensities. The
result is consistent with the theoretical ZnFe$_2$O$_4$(311) pole
figure viewed along [111] with rotation twins. The in-plane
orientation relationship is
ZnFe$_2$O$_4$[110]$\parallel$ZnO[11$\underline{2}$0]. Fig.
\ref{fig:Fig3ZnFeO}(b) shows the schematics for the
crystallographical orientation of ZnFe$_2$O$_4$ onto ZnO. Due to
the fcc structure of ZnFe$_2$O$_4$ (a=0.844 nm), it is not
difficult to understand its epitaxy onto hcp-ZnO (a=0.325 nm) with
twin-crystallites of ZnFe$_2$O$_4$ of an in-plane rotation by
60\degr. The coherence length of crystallites is around 20 nm in
the out-of-plane direction. The in-plane coherence length is
evaluated to be also as large as 20 nm by measuring the
diffraction of (311) at $\chi$$\sim$80\degr (not shown), nearly
parallel with the surface \cite{heinke:2145}. Due to the fcc
structure of ZnFe$_2$O$_4$ (a=0.844 nm), it is not difficult to
understand its crystallographical orientation onto hcp-ZnO
(a=0.325 nm) with twin-crystallites of ZnFe$_2$O$_4$ of an
in-plane rotation by 60\degr. The lattice mismatch between
ZnFe$_2$O$_4$ and ZnO is ~6\%.

\subsubsection{NiFe$_2$O$_4$ and NiFe$_2$O$_4$}

Generally, spinel ferrites (MFe$_2$O$_4$, M=Ni, Co, Fe, Mn, Zn)
have a large variety of magnetic properties and have significant
potential application in millimeter wave integrated circuitry and
magnetic recording \cite{suzuki01}. We also synthesized
NiFe$_2$O$_4$ and CoFe$_2$O$_4$ by (Ni, Fe) and (Co, Fe)
co-implantation, respectively \cite{zhou_ferrite}. The charge
states of Ni, Fe and Co, as well as the phase formation of spinel
have been confirmed by x-ray absorption. Nanocrystalline
NiFe$_2$O$_4$ and CoFe$_2$O$_4$ are as soft and hard magnet,
respectively, embedded inside ZnO. In the view of lattice
mismatch, our results suggest the epitaxy of spinel ferrites onto
ZnO, and even a multi-layered MFe$_2$O$_4$/ZnO structure given the
growth method compatibility by pulsed laser deposition or
molecular beam epitaxy for both materials \cite{suzuki01,ozgur05}.
Thus, a hybrid structure of spinel ferrites/semiconducting ZnO
could be a potential candidate for magneto-electronics devices.

\begin{figure} \center
\includegraphics[scale=0.8]{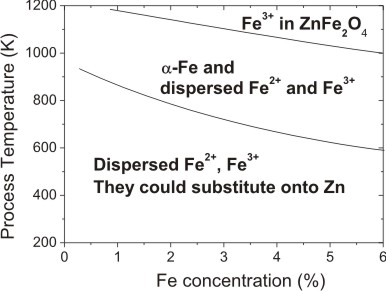}
\caption{The phase diagram of Fe in ZnO bulk crystals derived from
the data presented in this work. The process temperature refers to
the implantation or annealing temperature. Adapted from Ref.
\cite{zhouFe}.}\label{fig:phase_diagram}
\end{figure}

Summarizing our study on Fe implanted ZnO single crystals we
present a phase diagram of Fe in ZnO depending on the implantation
parameters, \ie~fluence, energy, and annealing temperature as
shown in Figure \ref{fig:phase_diagram}. Note that the materials
studied in this research are ZnO bulk crystals grown by
hydro-thermal method. They are semi-insulating in the as-purchased
state with n-type carrier concentration of 10$^{12}$-10$^{14}$
cm$^{-3}$. The phase diagram will likely be different for
epitaxial-ZnO and for p-type ZnO.

\subsection{Hidden secondary phases}
\begin{figure} \center
\includegraphics[scale=0.8]{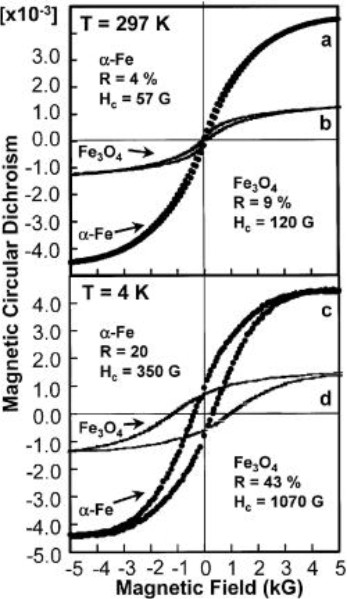}
\caption{MCD results showing the hysteretic response of magnetic
-Fe and Fe$_3$O$_4$ precipitates in a single crystal YSZ host.
Curves a and c show the MCD results for Fe at 297 and 4 K,
respectively, while curves b and d show the results for
Fe$_3$O$_4$ at these temperatures. The coercivity, H$_C$, and a
factor, R, representing the ratio of the remanent response at H=0
to the saturation value at high H, are indicated. Adapted from
Ref. \cite{honda:711}.}\label{fig:honda}
\end{figure}

"Hidden" ferromagnetic secondary phases, i.e. CoZn clusters
involving metallic Co have been found recently by Kaspar et al. in
epitaxial Co doped ZnO \cite{kaspar:201303}. Thus, they identified
the origin of ferromagnetism appearing after annealing the samples
in Zn vapor. Conventional characterization techniques indicate no
change after treatment. Element specific methods like x-ray
absorption fine structure are necessary to identify the secondary
phase as ferromagnetic. CoZn may form during deposition or
postgrowth processing under low-oxygen or vacuum conditions, even
in the absence of Zn vapor. They say: "CoZn is a particularly
insidious ferromagnetic secondary phase, since its low moment per
Co and low Curie temperature are in line with the expected
properties of ferromagnetic Co:ZnO. Thus, in the absence of
careful materials characterization, undetected CoZn could mimic
intrinsic ferromagnetism in Co:ZnO." Note that the change of
ferromagnetic properties in oxides containing metallic TM clusters
has been investigated much earlier for Fe clusters in YSZ
\cite{honda:711}. The Fe within those clusters changes from
metallic to ionic, dependent which annealing atmosphere, i.e.
reducing or oxidizing, is applied (Fig. \ref{fig:honda}).

\begin{figure} \center
\includegraphics[scale=0.6]{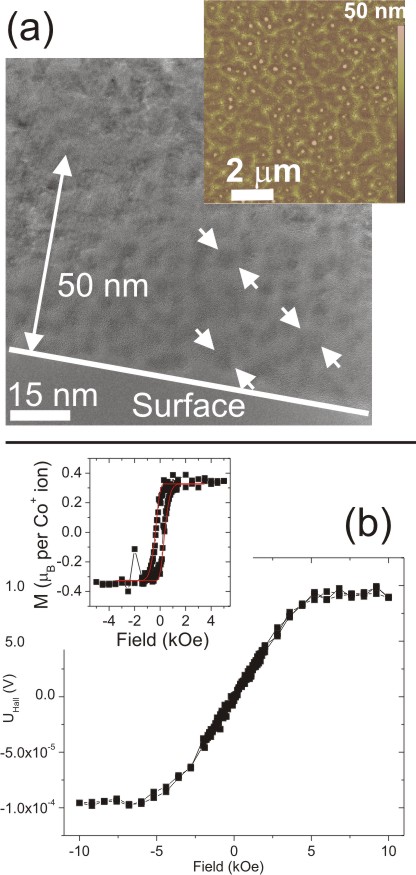}
\caption{(a) TEM micrograph of the near surface region of a
ZnO(0001) single crystal implanted with Co. The white arrows
indicate small crystalline clusters on the amporphous background.
The inset shows the surface topography measured using AFM. (b) AHE
at 5 K measured in perpendicular geometry with respect to the
magnetic field. The inset shows a magnetic hysteresis loop
measured at 5 K in parallel geometry. Adapted from Ref.
\cite{potzger:232504}.}\label{fig:disorder}
\end{figure}

One reason for "hidden" ferromagnetic clusters is their low
crystallinity in the sense of reduced long range crystalline order
or even amorphous phases. Such phases can occur for both metallic
or oxide clusters of TM. The structural disorder can even
influence the magnetic properties. As found during our research
\cite{potzger:232504}, such regions are created during
implantation of Co$^+$ ions into ZnO(0001) crystals located at a
liquid-nitrogen cooled target. The fluence was
1.6$\times$10$^{17}$ cm$^{-2}$ at an energy of 80 keV
corresponding to an atomic concentration of 25\%. At such large
fluence, superparamagnetic clusters are formed which cannot be
detected by the common XRD-scans and also can be easily overlooked
in TEM due to the dominant crystalline background from the
Wurtzite ZnO. Fig. \ref{fig:disorder} shows a compilation of those
results. Fig. \ref{fig:disorder}(a) shows a TEM micrograph of an
amorphousd dominated near-surface region with small crystalline
inclusions (white arrows). The upper right inset shows an AFM
micrograph of the surface of the sample. The  latter displays
regular hillock-like structures created by the implantation. Fig.
\ref{fig:disorder}(b) shows the Hall measurement at 5 K in
perpendicular geometry yielding anomalous Hall effect (AHE). SQUID
magnetometry (inset) shows a saturation magnetization of 0.35
$\mu_B$ per Co ion implanted. Recent element-specific measurements
using XMCD show that the ferromagnetic order mainly originates
from metallic Co with an admixture of ordered ionic Co.

\section{Summary, conclusions and comparison with other publications}The main message of this article is the following: There are
manifold sources for - especially weak - ferromagnetic hysteresis
in TM implanted ZnO single crystals. Ferromagnetic DMS has not
been observed.

Nanoscale clusters of intrinsic defects containing
(antiferromagnetically) interacting spins which are uncompensated
at the surface can easily be confused with regions of
ferromagnetic DMS, especially due to non-monotonic dependence on
the doping concentration. In part 4 we showed strong evidence that
a combination of microscopic defects in ZnO with delicate point
like defects, both created upon energy impact, leads to weak
ferromagnetic properties.  In some cases, the ferromagnetic
properties degrade or disappear after exposure to ambient
conditions for some days or due to annealing at temperatures in
the range of 1073 K. We would like to again point out that energy
impact on low quality ZnO yields larger decomposition tendency
than on high quality ZnO single crystals. In that way, nanoscale
defect clusters can be formed easily in the pre-annealed samples
and either reoxidize or annihilate with microscopic defects after
annealing at large temperatures. The tiny defect responsible for
the ferromagnetic properties remains unidentified. Intuitively,
defects resulting from O-excess are not easily acceptable. On the
other hand, the above preparation techniques decomposing ZnO can
leave regions with Zn- and O-excess behind, even if a net oxygen
loss occurs. The assumption of oxygen p-states to be the origin of
the ferromagnetism would suggest defect magnetism to be a rather
universal phenomenon in oxides. We do not want to completely
exclude an influence of spurious TM contamination like Fe or Cu.
Especially, since they are also attracted by the microscopic
defects, i.e. the local concentration of TM contaminants might be
much larger than for the whole sample. However, the above
considerations make a scenario involving intrinsic defects of ZnO
most likely.

In part 5 of this article, we presented metallic cluster formation
in Fe, Ni and Co implanted ZnO single crystals. A general feature
of such clusters is superparamagnetism, i.e. reduced critical
temperatures, making the identification the the materials via the
Curie temperature impossible. A correlation between magnetometry
as well as high-resolution structural analysis helps, e.g., to
understand the appearance of magnetic out-of-plane anisotropy in
Co:ZnO and the XRD-invisibility of metallic Fe in Fe:ZnO. At mild
implantation temperatures, the majority of the implanted ions,
however, is in ionic state. Those ions are paramagnetic exhibiting
Ni$^{2+}$, Co$^{2+}$, and mixed Fe$^{2+/3+}$ oxidation states.
Surprisingly, ZnO implanted with Co at low temperatures exhibits a
huge amount of Co substitutional sites in the Wurtzite lattice
\cite{Potzger_Co}. After implantation of the pre-annealed crystals
and post-annealing \cite{potzger:062107} no crystalline secondary
phases could be detected for Fe or Ni implanted samples. The
suppression of the cluster formation can be explained by
microscopic defects created by the pre-annealing. These can act as
sinks for the implanted TM ions where amorphous non-magnetic
Zn-TM-O complexes are formed. Note that energy impact on low
quality ZnO yields larger decomposition tendency than on high
quality ZnO single crystals \cite{coleman:231912} which might
promote the formation of such complexes during post-annealing.
Large local concentration of TM ions achieved either by large
fluences implanted or large annealing temperatures leads to the
formation of inverted spinels. Those are ferrimagnetic and also
lead to magnetic hysteresis loops. Besides crystalline
ferro(i)magnetic clusters, low crystallinity in the sense of
reduced long range structural order, or even amorphous regions
have been found. Those are generally not detectable by common XRD
scans.

Comparing our results with other groups, we realized that there is
unfortunately only limited interest in the investigation of
secondary phase formation in TM doped ZnO. The pioneering work on
Co implanted ZnO revealed metallic Co clusters to be the main
reason for ferromagnetism \cite{norton03}. We extended these
investigations to metallic Fe and Ni clusters. Inverted
ZnTM$_2$O$_4$ spinel materials are well known for some decades.
They are prepared intentionally for various technological
applications. In spintronics, the interest in inverted spinel
ZnTM$_2$O$_4$ rises due to its simultaneous ferromagnetic and
semiconducting properties. One example is ZnCo$_2$O$_4$ with
magnetic and semiconducting properties \cite{kim:7387}. On the
other hand, Co ions in structurally perfect films of the model
system Zn$_{1-x}$Co$_x$O without second phases are proven to be
paramagnetic \cite{ney:157201}. These observations correspond to
our results, although disorder has been introduced by the ion
implantation. Recently it has been shown that a serious source for
ferromagnetic properties of nanoparticles is uncompensated surface
spins. Such spins can occur for any (antiferromagnetically)
interacting paramagnetic electrons such as d electrons from TM and
unpaired p-like electrons from intrinsic defects. The former has
been discussed for ZnO containing CuO-clusters
\cite{sudakar:054423}. They are also known for pure CoO
\cite{dutta08} or NiO \cite{yi:224402} nanoparticles. Note that
large saturation magnetization has been achieved for CoO
nanoparticles in ref. \cite{dutta08}. Uncompensated surface spins
are nearly ruled out as the only source for the observed
room-temperature ferromagnetism. Defect induced ferromagnetism, as
in our case, has been observed in Ar$^+$ irradiated ZnO
\cite{borgesArZnO}. There is only limited potential for
application of defect induced ferromagnetism. Suggestions to
overcome the superparamagnetic limit in magnetic recording media
was made by Hernando et al. \cite{hernando:052403} since defect
induced ferromagnetism appears to show different temperature
dependence as compared to the conventional one.

The example of ferromagnetic GaMnAs, however, proofs that our
results do not exclude the existence of ZnO based DMS. Instead,
more experimental work has to be done while monitoring unwanted
magnetic properties by appropriate analysis methods. Alternative
materials to TM diluted ZnO are TM based ferromagnetic spinels.
The rich family of those spinels offers  wide possibilities of
tuning the bandgap, electronic conduction parameters and magnetic
properties by selecting appropriate TM and inversion grade.

\begin{acknowledgement}
We would lime to thank E. Christalle, M. Missbach, A. Kunz, A.
Scholz, J. Kreher, C. Neisser, H. Felsmann, as well as J.
Schneider, S. Eisenwinder, and F. Ludewig for technical support.
For scientific support as well as fruitful discussions we would
like to thank M. Helm and W. M\"{o}ller, J. Fassbender, W.
Skorupa, A. M\"{u}cklich, R. Gr\"{o}tzschel, A. Shalimov, J.
Grenzer, K. Kuepper, H. Schmidt, B. Schmidt, G. Brauer, H.
Reuther, C. B\"{a}htz, F. Eichhorn, V. Cantelli, T.
Hermannsd\"{o}rfer, T. Papageorgiou, Q. Xu, A. Janotti, M. Lorenz,
E. Arenholz, J. D. Denlinger, S. Gemming, N. Volbers, J. Sann, S.
Lautenschl\"{a}ger, W. Anwandt, G. Talut, G. Zhang.

The author (S.Z.) thanks for the financial funding from the
Bundesministerium f\"{u}r Bildung und Forschung (FKZ03N8708).
\end{acknowledgement}

%

\providecommand{\WileyBibTextsc}{}
\let\textsc\WileyBibTextsc
\providecommand{\othercit}{} \providecommand{\jr}[1]{#1}
\providecommand{\etal}{~et~al.}

\end{document}